\documentclass{article}
\usepackage{graphicx}  
\usepackage{amsmath}   
\usepackage[compress]{cite}
\usepackage{amssymb}   
\usepackage{bm} 
\usepackage{dcolumn}
\usepackage{color}
\usepackage{mathrsfs}
\usepackage{gensymb}
\usepackage{float}
\usepackage{amsfonts}
\usepackage{varioref}
\usepackage{textcomp}
\usepackage[caption=true]{subfig}

\RequirePackage[colorlinks,citecolor=blue,urlcolor=magenta,linkcolor=blue]{hyperref}
\allowdisplaybreaks
\def\gr{general relativity}

\labelformat{section}{Section #1} 
\labelformat{subsection}{Section #1} 
\labelformat{subsubsection}{Section #1}
\labelformat{subsubsubsection}{Section #1}
\labelformat{equation}{Eq.~(#1)} 
\labelformat{figure}{Fig.~#1} 
\labelformat{subfigure}{Fig.~\thefigure#1} 
\labelformat{table}{Tab.~#1} 
\labelformat{appendix}{Appendix #1}
\title{Imprints of Einstein-Maxwell dilaton-axion gravity in the observed shadows of Sgr A* and M87* }
\author{Siddharth Kumar Sahoo\footnote{521ph1007@nitrkl.ac.in}~,
Neeraj Yadav\footnote{418ph5062@nitrkl.ac.in}~
and
Indrani Banerjee\footnote{banerjeein@nitrkl.ac.in}\\
{\small{\hspace{-1.5cm}Department of Physics and Astronomy, National Institute of Technology, Rourkela, Odisha-769008 India}}}

\date{ }  
\begin{document}
  
\maketitle
\begin{abstract}
Einstein-Maxwell dilaton-axion (EMDA) gravity provides a simple framework to investigate the signatures of string theory. The axion and the dilaton fields arising in EMDA gravity have important implications in inflationary cosmology and in addressing the late time acceleration of the universe. It is therefore instructive to explore the implications of such a model in explaining the astrophysical observations. In this work we explore the role of EMDA gravity in explaining the observed shadows of black holes (M87* and Sgr A*) released by the Event Horizon Telescope (EHT) collaboration. The Kerr-Sen metric represents the exact, stationary and axisymmetric black hole solution of EMDA gravity. Such a black hole is characterized by the angular momentum $a$ acquired from the axionic field and the dilatonic charge $r_2$ arising from string compactifications. We study the role of spin and the dilaton charge in modifying the shape and size of the black hole shadow. We note that black holes with larger dilaton charge cast a smaller shadow. We investigate the consequences of such a result in addressing the EHT observations of M87* and Sgr A*. Our analysis reveals that the shadow of M87* exhibits a preference towards the Kerr scenario. However, when 10\% offset in the shadow diameter is considered, $0.1\lesssim r_2\lesssim 0.3$ is observationally favored within 1-$\sigma$. The shadow of Sgr A* on the other hand shows a preference towards the Kerr-Sen scenario since the central value of its shadow can be better explained by a non-zero dilaton charge $0.1 \lesssim r_2 \lesssim 0.4$. However, when the 1-$\sigma$ interval is considered the Kerr scenario is included. We discuss the implications of our results.

\end{abstract}
\section{Introduction}
\label{sec:intro}
General relativity (GR), the successor of Newtonian theory of gravity has radically changed our understanding pertaining to gravitational interaction. In GR the mass of a body produces curvature in the spacetime, which changes the metric of the spacetime from Minkowski metric \cite{Weinberg:1972kfs}. The particles in curved spacetime move along geodesics which are obtained by solving the geodesic equation associated with the metric describing the spacetime. The metric itself is obtained by solving the Einstein field equations and depends crucially on the matter distribution. GR has many interesting predictions \cite{Weinberg:1972kfs}, namely, the perihelion precession of mercury, the bending of light, the gravitational redshift of radiation from distant stars, to name a few, which have been experimentally verified \cite{Will:2014kxa,Will:1993ns}. The detection of gravitational waves by the LIGO-VIRGO collaboration \cite{LIGOScientific:2016aoc,LIGOScientific:2017vwq,LIGOScientific:2018mvr} and the release of black hole images of M87* and Sgr A* by the Event Horizon Telescope collaboration \cite{EventHorizonTelescope:2019dse,EventHorizonTelescope:2019ggy,EventHorizonTelescope:2019pgp,EventHorizonTelescope:2019uob,EventHorizonTelescope:2019jan,EventHorizonTelescope:2019ths,EventHorizonTelescope:2022wkp,EventHorizonTelescope:2022apq,EventHorizonTelescope:2022wok,EventHorizonTelescope:2022exc,EventHorizonTelescope:2022urf,EventHorizonTelescope:2022xqj} have further demonstrated GR as a successful theory of gravity.\\

Despite being a very successful theory, GR also has certain limitations. The theory allows formation of singularities  \cite{Penrose:1964wq,Hawking:1976ra,Hawking:1973uf} namely, the black hole and the big bang singularities,
where the theory loses its predictive power \cite{Wald:1984rg,Penrose:1964wq,Hawking:1976ra,Christodoulou:1991yfa}. This indicates that GR is not a complete theory of gravity \cite{Will:2014kxa} and that at very small length scales it must receive considerable corrections from a more complete theory that incorporates its quantum nature\cite{Rovelli:1996dv,Dowker:2005tz,Ashtekar:2006rx,Kothawala:2013maa}  \cite{Wigner1997}. In the observational front GR falls short in explaining the nature of dark matter\cite{Milgrom:1983pn,Bekenstein:1984tv,Milgrom:2003ui} and dark energy\cite{Clifton:2011jh,SupernovaCosmologyProject:1998vns,SupernovaSearchTeam:1998fmf}, which are invoked to explain the flat rotation curves of galaxies and the accelerated expansion of the universe, respectively. 
These inadequacies have lead to the development of many alternate theories of gravity which address the limitations of GR \cite{Sotiriou:2008rp,DeFelice:2010aj,Capozziello:2011et,Maartens:2003tw,Overduin:1997sri,Kobayashi:2019hrl,Hammond:2002rm} and deviate from GR in the strong field regime. Therefore, to test the effectiveness of alternative theories of gravity, it is necessary to study how effectively they explain observations related to strong field tests of gravity \cite{Berti:2015itd,Psaltis:2008bb}.\\

The alternatives to GR include higher curvature gravity, e.g., $f(R)$ gravity,\cite{Nojiri:2003ft,Nojiri:2006gh,Capozziello:2006dj}  and Lanczos Lovelock models\cite{Lanczos:1932zz,Lanczos:1938sf,lovelock1971einstein,Padmanabhan:2013xyr}, extra dimensional models \cite{Shiromizu:1999wj,Dadhich:2000am,Harko:2004ui,Carames:2012gr,Kobayashi:2006jw,Chakraborty:2014xla,Chakraborty:2015bja} and scalar-tensor/scalar-vector-tensor theories of gravity\cite{Horndeski:1974wa,Sotiriou:2013qea,Babichev:2016rlq,Charmousis:2015txa,Bhattacharya:2016naa}. Many of these models are string inspired which provides a framework for force unification\cite{Horava:1995qa,Horava:1996ma,Polchinski:1998rq,Polchinski:1998rr} . In this work we intend to discern the signatures of the string inspired model, namely, the Einstein-Maxwell dilaton-axion (EMDA) gravity, from observations related to black hole shadows. EMDA, a scalar-vector-tensor theory of gravity, arises in the low energy effective action of superstring theories\cite{Sen:1992ua} on compactifying the ten dimensional heterotic string theory on a six dimensional torus. In the EMDA theory the scalar field dilaton and the pseudo scalar axion are coupled to the Maxwell field and the metric. The axion and dilaton fields which originate from string compactifications have interesting implications in inflationary cosmology and the late-time acceleration of the universe \cite{Sonner:2006yn,Catena:2007jf}. It is, therefore, important to explore the footprints of EMDA gravity in astrophysical observations which is the goal of the present work. 
In particular, we aim to decipher the imprints of the dilaton charge in black holes from observations related to black hole shadows.

Black holes(BH) are compact objects with extremely strong gravity. Among the various systems that possess strong gravitational field, black holes are the most interesting and the simplest ones. Different black hole solutions have been constructed in the context of string inspired low-energy effective theories\cite{Gibbons:1987ps,Garfinkle:1990qj,Horowitz:1991cd,Kallosh:1993yg}. Interestingly, the charge neutral axisymmetric black hole solution in string theory resembles the Kerr solution in GR \cite{Thorne:1986iy,Psaltis:2007cw}. In EMDA gravity the stationary and axisymmetric black hole solution is represented by the Kerr-Sen metric which is similar to the Kerr-Newman spacetime in GR. Despite the similarities, the intrinsic geometry of the two black holes vary considerably which have been explored extensively in the past \cite{Hioki:2008zw,Pradhan:2015yea,Guo:2019lur,Uniyal:2017yll}.
Investigating the observational signatures of the Kerr-Sen black hole is important as it can provide an indirect testbed for string theory. Astrophysical signatures of Kerr-Sen black hole have been studied previously in the context of photon motion, null geodesics, strong gravitational lensing and black hole shadows \cite{Gyulchev:2006zg,An:2017hby,Younsi:2016azx,Hioki:2008zw,Mizuno:2018lxz,Xavier:2020egv}. In \cite{Xavier:2020egv}, the authors have worked out the shadow of the Kerr-Sen black hole but there they have not compared their result with the observed shadows and hence no constrain on the dilaton charge was reported. Recently, the shadow of dyonic Kerr-Sen black holes have been studied \cite{Jana:2023sil} and an upper bound on the magnetic monopole charge of Sgr A* has been mentioned. 
We explore the role of the dilaton charge in modifying the structure of the black hole shadow from that of the Kerr scenario. We compare the theoretically derived black hole shadow (which depends on the dilaton charge and the spin) with that of the observed images of M87* and Sgr A* (released by the EHT collaboration). 
Such a study enables us to establish constrains on the dilaton parameter of the Kerr-Sen black hole and allows us to comment on the possible feasibility of string theory in explaining the observed black hole shadows. 

The structure of the paper is as follows: In \ref{Sec2} we give a brief overview of the Kerr-Sen BH.  In \ref{Sec3} we derive the shadow outline of the Kerr-Sen BH.  In \ref{Sec4} we discuss our results related to constrains on the dilaton parameter \(r_{2}\) from EHT observations of M87* and SgrA*. We give summary of our results and concluding remarks in Section \ref{Sec6}.  In our paper we have chosen the metric signature \((-,+,+,+)\) and used geometrized units \(G=c=1\).

\section{Black hole in Einstein-Maxwell dilaton axion gravity}\label{Sec2}
The Einstein-Maxwell dilaton-axion (EMDA) gravity \cite{Sen:1992ua,Rogatko:2002qe} results from the compactification of ten dimensional heterotic string theory on a six dimensional torus $T^6$. In EMDA gravity, $N=4$, $d=4$ supergravity is coupled to $N=4$ super Yang-Mills theory which can be suitably truncated to a pure supergravity theory exhibiting $S$ and $T$ dualities. The bosonic sector of this supergravity theory when coupled to the $U(1)$ gauge field is known as the Einstein-Maxwell dilaton-axion (EMDA) gravity \cite{Rogatko:2002qe} which provides a simple framework to study classical solutions. The four dimensional effective action for EMDA gravity consists of a generalization of the Einstein-Maxwell action such that the metric $g_{\mu\nu}$ is coupled to the dilaton field $\chi$, the $U(1)$ gauge field $A_\mu$ and the Kalb-Ramond field strength tensor ${H}_{\alpha\beta\gamma}$. The action corresponding to EMDA gravity assumes the form,
\begin{align}
S=\frac{1}{16\pi} \int \sqrt{-g}d^{4}x(R-2\partial_{\mu} \chi \partial^{\mu} \chi-\frac{1}{3} H_{\rho \sigma \delta}H^{\rho \sigma \delta}+ e^{-2 \chi} F_{\alpha \beta} F^{\alpha \beta})
\label{S2-1}
\end{align}
In \ref{S2-1} $g$ is the determinant and $R$ the Ricci scalar associated with the 4-dimensional metric $g_{\mu\nu}$, 
\(\chi\) represents the dilatonic field, \(F_{\mu \nu}=\nabla_{\mu} A_{\nu}-\nabla_{\nu} A_{\mu}\) is Maxwell field strength tensor and \(H_{\rho \sigma \delta}\) is given by
\begin{equation}
\begin{split}
H_{\rho \sigma \delta}=\nabla_{\rho} B_{\sigma \delta}+\nabla_{\sigma} B_{\delta \rho}+\nabla_{\delta} B_{\rho \sigma}\\-A_{\rho}B_{\sigma \delta}-A_{\sigma}B_{\delta \rho}-A_{\delta}B_{\rho \sigma}
\end{split}
\label{S2-2}
\end{equation}
where \(A_{\mu}\) is the vector potential and \(B_{\mu \nu}\) is the second rank antisymmetric tensor field called the Kalb-Ramond field while its cyclic permutation with $A_\mu$ denotes the Chern-Simons term. In four dimensions the Kalb-Ramond field strength tensor \(H_{\rho \sigma \delta}\) can be written in terms of the pseudo-scalar axion field $\psi$, such that,
\begin{align}
\label{S2-3}
{H}_{\alpha\beta\delta} = \frac{1}{2}e^{4\chi}\epsilon_{\alpha\beta\delta\gamma}\partial^{\gamma}\psi
\end{align}
The action in \ref{S2-1} written in terms of the axion field assumes the form,
\begin{align}
\label{S2-4}
S = \frac{1}{16\pi}\int\sqrt{-g}~d^{4}x\bigg{[}{R} - 2\partial_{\nu}\chi\partial^{\nu}\chi - \frac{1}{2}e^{4\chi}\partial_{\nu}\psi\partial^{\nu}\psi + e^{-2\chi}{F}_{\rho\sigma}{F}^{\rho\sigma}  + \psi{F}_{\rho\sigma}\tilde{{F}}^{\rho\sigma}\bigg{]} 
\end{align}
Variation of the action with respect to the dilaton, axion and Maxwell fields give their corresponding equations of motion. The equation of motion associated with the axion field is given by,
\begin{align}
\nabla_{\mu}\nabla^{\mu}\psi + 4\nabla_{\nu}\psi\nabla^{\nu}\psi - e^{-4\chi}{F}_{\rho\sigma}\tilde{{F}}^{\rho\sigma} &= 0 \label{S2-5}
\end{align} 
while that of the dilaton field assumes the form,
\begin{align}
\nabla_{\mu}\nabla^{\mu}\chi - \frac{1}{2}e^{4\chi}\nabla_{\mu}\psi\nabla^{\mu}\psi + \frac{1}{2}e^{-2\chi}{F}^{2} &= 0
\label{S2-6}
\end{align}
The Maxwell's equations with couplings to the dilaton and the axion fields are given by,
\begin{align}
\nabla_{\mu}(e^{-2\chi}{F}^{\mu\nu} + \psi\tilde{{F}}^{\mu\nu}) &= 0,\label{S2-7}\\
\nabla_{\mu}(\tilde{{F}}^{\mu\nu}) &= 0 \label{S2-8}
\end{align}
Solving the aforesaid equations one obtains solutions for the dilaton, axion and the Maxwell field, respectively \cite{Ganguly:2014pwa,Sen:1992ua,Rogatko:2002qe},
\begin{align}
e^{2\chi} &= \frac{r^{2} + a^{2}\cos^{2}\theta}{r(r + r_{2}) + a^{2}\cos^{2}\theta}\label{S2-9}\\
\psi &= \frac{q^{2}}{{M}}\frac{a\cos\theta}{r^{2} + a^{2}\cos^{2}\theta} \label{S2-10}\\
A & =\frac{qr}{\tilde{\Sigma}}\bigg(-dt +a \mathrm{sin}^2\theta d\phi\bigg)
\label{S2-11}
\end{align}
where $M$ is the mass, $a$ is the spin and $q$ is the charge of the black hole. In \ref{S2-9} $r_2$ is associated with the dilaton parameter and is given by $r_{2} = \frac{q^{2}}{{M}}e^{2\chi_{0}}$ where $\chi_{0}$ represents the asymptotic value of the dilatonic field. The dilaton parameter also depends on the electric charge of the black hole, which owes its origin from the axion-photon coupling and not from the in-falling charged particles. This is because the axion and dilaton field strengths vanish if the electric charge $q=0$ (see \ref{S2-10} and \ref{S2-9}). It is further important to note that the axion field renders a non-zero spin to the black hole since the field strength corresponding to the axion field vanishes if the black hole is non-rotating (\ref{S2-10}).

Varying the action with respect to the metric gives the Einstein field equations, 
\begin{align}
{G}_{\mu\nu} = {T}_{\mu\nu}({F},\chi,\psi) \label{S2-12}
\end{align}
where, ${G}_{\mu\nu}$ is the Einstein tensor and ${T}_{\mu\nu}$ the energy-momentum tensor which is given by,
\begin{align}
\label{S2-13}
{T}_{\mu\nu}({F},\chi,\psi)& = e^{2\chi}(4{F}_{\mu\rho}{F}_{\nu}^{\rho} - g_{\mu\nu}{F}^{2}) - g_{\mu\nu}(2\partial_{\gamma}\chi\partial^{\gamma}\chi + \frac{1}{2}e^{4\chi}\partial_{\gamma}\psi\partial^{\gamma}\psi) 
\nonumber \\
&+ \partial_{\mu}\chi\partial_{\nu}\chi + e^{4\chi}
\partial_{\mu}\psi\partial_{\nu}\psi
\end{align}
The Kerr-Sen metric \cite{Sen:1992ua} is obtained when one looks for the stationary and axisymmetric solution of the aforesaid Einstein’s equations \cite{Garcia:1995qz,Ghezelbash:2012qn,Bernard:2016wqo}. In Boyer-Lindquist coordinates the Kerr-Sen metric takes the form,
\begin{align}
\label{S2-14}
ds^{2} &= - \bigg{(}1 - \frac{2{M}r}{\tilde{\rho}}\bigg{)}~dt^{2} + \frac{\tilde{\rho}}{\Delta}(dr^{2} + \Delta d\theta^{2}) - \frac{4a{M}r}{\tilde{\rho}}\sin^{2}\theta dt d\phi  \nonumber \\
&+ \sin^{2}\theta d\phi^{2}\bigg{[}r(r+r_{2}) + a^{2} + \frac{2 {M}ra^{2}\sin^{2}\theta}{\tilde{\rho}}\bigg{]}
\end{align}
where,
\begin{align}
\label{S2-14a}
\tilde{\rho} &= r(r + r_{2}) + a^{2}\cos^{2}\theta \tag {14a}\nonumber\\
\Delta &= r(r + r_{2}) - 2{M}r + a^{2} \tag {14b}\nonumber
\end{align}
The non-rotating counterpart of the Kerr-Sen metric corresponds to a pure dilaton black hole characterized by its mass, electric charge and asymptotic value of the dilaton field \cite{Garfinkle:1990qj,Yazadjiev:1999xq}.

In order to obtain the event horizon $r_h$ of the Kerr-Sen black hole one solves for $g^{rr}=\Delta=0$, which gives,
\begin{align}
\label{S2-15}
r_h={M}-\frac{r_2}{2} +\sqrt{\bigg({M}-\frac{r_2}{2}\bigg)^2 - a^2}
\end{align}
Since $r_{2} = \frac{q^{2}}{{M}}e^{2\chi_{0}}>0$, the presence of real, positive event horizon requires $0\leq \frac{r_2}{{M}} \leq 2$ (see \ref{S2-15}). Since we are interested in black hole solutions we will be interested in this regime of $r_2$ in this work. 
\section{Shadow of Kerr-Sen black holes}\label{Sec3}
When photons from a distant astrophysical object or the accretion disk surrounding the black hole come close to the black hole horizon, a few of them get trapped inside the horizon while others escape to infinity. Since some photons get trapped inside the horizon, the observer sees a dark patch in the image of the black hole, known as the black hole shadow. The outline of the black hole shadow is associated with the motion of photons near the event horizon and hence we expect to extract valuable information regarding the nature of strong gravity from the shape and size of the black hole shadow \cite{Gralla:2019xty,Bambi:2019tjh,Hioki:2009na,Vagnozzi:2019apd,Banerjee:2019nnj}. It may be noted that the shape of the shadow depends on the background spacetime while the size of the shadow is related to the mass and distance as well as the background metric. Thus, a non-rotating black hole gives rise to a circular shadow in which case the size is the only parameter based on which one can study deviations from the Schwarzschild geometry in GR \cite{Cunha:2018acu,Vries_1999}. Rotating black holes cast a non-circular shadow provided the black hole is viewed at a high inclination angle. In such a scenario, both the size and the shape of the shadow can be used to study deviation from GR \cite{Cunha:2018gql,Gralla:2019xty,Bambi:2019tjh,Hioki:2009na,Vagnozzi:2019apd,Banerjee:2019nnj,Mizuno:2018lxz,Roy:2019esk}. \\
In this section we investigate the motion of photons in the Kerr-Sen background. This enables us to compute the outline of the black hole shadow in EMDA gravity which in turn can be compared with the Kerr scenario in \gr. 

For a stationary, axisymmetric metric, the Lagrangian $\mathcal{L}$ for the motion of any test particle is given by,
\begin{equation}\label{eq:16}
\mathcal{L}(x^{\mu},\dot{x^\mu})=\frac{1}{2}g_{\mu\nu} \dot{x}^{\mu} \dot{x}^{\nu}=\frac{1}{2}\left(g_{tt}\dot{t}^{2}+g_{rr}\dot{r}^{2}+g_{\theta\theta}\dot{\theta}^{2}+g_{\phi\phi}\dot{\phi}^{2}+2g_{t\phi}\dot{t}\dot{\phi}\right) 
\end{equation}
The action  $\mathcal{S}$ representing the motion of test particles satisfying the Hamilton-Jacobi equation is given by,
\begin{equation}\label{eq:17}
\mathcal{H}(x^{\nu},p_{\nu})+\frac{\partial{\mathcal{S}}}{\partial{\lambda}}=0 
\end{equation}
where $\mathcal{H}$ is the Hamiltonian, $\lambda$ is a curve parameter, and $p_\mu$, the conjugate momentum corresponding to the coordinate $x^{\mu}$ is
\begin{equation}\label{eq:19}
p_{\mu}=\frac{\partial{\mathcal{S}}}{\partial{x^{\mu}}}=\frac{\partial{\mathcal{L}}}{\partial{\dot{x}^{\mu}}}=g_{\mu\nu}\dot{x}^{\nu}
\end{equation}
The Hamiltonian is given by,
\begin{equation}\label{eq:18}
\mathcal{H}(x^{\nu},p_{\nu})=\frac{1}{2}g^{\mu\nu}p_{\mu}p_{\nu}=\frac{k}{2}=0
\end{equation}
where k denotes the rest mass of the test particle which is zero for photons.
Since the Kerr-Sen metric does not explicitly depend on t and $\phi$, the first term in the the Euler-Lagrange equation
\begin{equation}\label{eq:23}
\frac{\partial{\mathcal{L}}}{\partial{x^{\mu}}}-\frac{d}{d\lambda}\left(\frac{\partial{\mathcal{L}}}{\partial{\dot{x^{\mu}}}}\right)=0
\end{equation}
is zero. Therefore, the energy E and the angular momentum $L_{z}$ of the photon are conserved. Using \ref{eq:19} these constants are given by
\begin{equation}\label{eq:24}
E=-g_{tt}\dot{t}-g_{t\phi}\dot{\phi}=-p_{t}, \hspace{1cm} L_{z}=g_{t\phi}\dot{t} + g_{\phi\phi}\dot{\phi}=p_{\phi}
\end{equation}
We further note from \ref{eq:19} that, 
\begin{equation}\label{eq:21}
p_t=\frac{\partial{\mathcal{S}}}{\partial{{t}}}=g_{tt}\dot{t}+g_{t\phi}\dot{\phi}=-E, \hspace{1cm}
p_\phi=\frac{\partial{\mathcal{S}}}{\partial{{\phi}}}=g_{t\phi}\dot{t}+g_{\phi\phi}\dot{\phi}=L_{z}
\end{equation}

\begin{equation}\label{eq:22}
p_r=\frac{\partial{\mathcal{S}}}{\partial{{r}}}=g_{rr}\dot{r}, \hspace{1cm}
p_\theta=\frac{\partial{\mathcal{S}}}{\partial{{\theta}}}=g_{\theta\theta}\dot{\theta}
\end{equation}
Integrating \ref{eq:21} the action $\mathcal{S}$ can be written as 
\begin{equation}\label{eq:25}
\mathcal{S}=-Et+L_{z}\phi+\bar{\mathcal{S}}(r,\theta)
\end{equation}
It turns out that $\mathcal{\bar{S}}(r,\theta)$ can be separated in $r$ and $\theta$ giving us
\begin{equation}\label{eq:26}
\mathcal{S}=-Et+L_{z}\phi+\mathcal{S}^{r}(r)+\mathcal{S}^{\theta}(\theta)
\end{equation}
From \ref{eq:18} we have $g^{\mu\nu}p_{\mu}p_{\nu}=0$ giving us
\begin{equation}\label{eq:20}
g^{tt}p_{t}^2+g^{rr}p_{r}^2+g^{\theta\theta}p_{\theta}^2+g^{\phi\phi}p_{\phi}^2+2g^{t\phi}p_{t}p_{\phi}=0
\end{equation}
Using \ref{eq:21}, \ref{eq:22} and \ref{eq:26}, \ref{eq:20} can be written as 
\begin{equation}\label{eq:27}
g^{tt}E^2-2g^{t\phi}E L_{z}+g^{\phi\phi}L_{z}^2+g^{rr}\left(\frac{d\mathcal{S}^r}{dr}\right)^2+g^{\theta\theta}\left(\frac{d\mathcal{S}^{\theta}}{d\theta}\right)^{2}=0
\end{equation}
which on substitution of the metric components $g^{\mu\nu}$ (see \ref{S2-14}) gives
\begin{equation}\label{eq:28}
\begin{split}
\left[\Delta a^2 \sin^2\theta-(r(r+r_{2})+a^2)^{2}\right]\frac{E^2}{\Delta}+\frac{4MraEL_{z}}{\Delta}+\frac{L_{z}^2}{\Delta \sin^2\theta}\left(\tilde{\rho}-2Mr\right)+\\ \Delta \left(\frac{d\mathcal{S}^r}{dr}\right)^2+\left(\frac{d\mathcal{S}^\theta}{d\theta}\right)^2=0
\end{split}
\end{equation}
The above equation can be separated in r and $\theta$ such that, 
\begin{equation}\label{eq:29}
\begin{split}
\Delta \left(\frac{d\mathcal{S}^r}{dr}\right)^2+a^2E^2+L_{z}^2-\frac{aL_{z}^2}{\Delta}-\frac{E^2}{\Delta}(r(r+r_{2})+a^2)^2+ \frac{4MraL_{z}E}{\Delta} =\\
-\left( \frac{d\mathcal{S}^\theta}{d\theta}\right)^2+aE^2\cos^2\theta-L_{z}^2\cot^2\theta=-Q
\end{split}
\end{equation}
where Q is called the Carter's constant.
From \ref{eq:29} the angular part is given by
\begin{equation}\label{eq:30}
\left( \frac{d\mathcal{S}^\theta}{d\theta}\right)=\sqrt{Q-L_{z}^2\cot^2\theta+a^2E^2\cos^2\theta}=\sqrt{\Theta(\theta})
\end{equation}
where 
\begin{equation}\label{eq:31}
Q-L_{z}^2\cot^2\theta+a^2E^2\cos^2\theta=\Theta(\theta)
\end{equation}
The radial equation is given by
\begin{equation}\label{eq:32}
\mathcal{V}(r)=\Delta^2\left(\frac{d\mathcal{S}^r}{dr}\right)^2
\end{equation}
where 
\begin{equation}\label{eq:33}
\mathcal{V}(r)=-Q\Delta-a^2E^2\Delta+E^2\left( r(r+r_{2})+a^2\right)^2+a^2L_{z}^2-\Delta L_{z}^2-4MraEL_{z}
\end{equation}
We also note that 
\begin{align}
\label{S3-1}
\dot{r}=p^r=g^{rr}\frac{d\mathcal{S}^r}{dr}=\frac{\Delta}{\tilde{\rho}}\frac{\sqrt{\mathcal{V}(r)}}{\Delta}
\end{align}
while 
\begin{align}
\label{S3-2}
\dot{\theta}=p^\theta=g^{\theta\theta}\frac{d\mathcal{S}^\theta}{d\theta}=\frac{\sqrt{\Theta(\theta)}}{\tilde{\rho}}
\end{align}
Therefore the first order geodesic equations for $r$ and $\theta$ can be respectively written as,
 \begin{align}
 \label{S3-3}
 \bigg(\frac{\tilde{\rho}}{E}\bigg)^2\dot{r}^2= a^2\xi^2 + (r(r+r_2) + a^2)^2 -4Mra\xi -\Delta (\eta + a^2 +\xi^2) ~~
 \rm and
 \end{align}
\begin{align}
\label{S3-4}
\bigg(\frac{\tilde{\rho}}{E}\bigg)^2\dot{\theta}^2=\eta  + a^2 \cos^2 \theta -\xi^2 \cot^2\theta
\end{align}
where $\xi=L_z/E$ and $\eta=Q/E^2$ represent the two impact parameters. While $\xi$ denotes the distance from the axis of rotation, $\eta$ signifies the distance from the equatorial plane.

The first order geodesic equations for $t$ and $\phi$ are obtained from \ref{eq:21} and are given by,
\begin{equation}
\label{S3-5}
\dot{t}=\frac{E[((r+r_{2})r+a^{2})^{2}-\Delta a^{2} \sin^{2}\theta]}{\tilde{\rho}\Delta}-\frac{2MarL_z}{\tilde{\rho}\Delta}
\end{equation}
\begin{equation}
\label{S3-6}
\dot{\phi}=\left(\frac{\tilde{\rho}-2Mr}{\tilde{\rho}\Delta}\right)\frac{L_z}{\sin^{2}\theta}+\frac{2MraE}{\tilde{\rho}\Delta}
\end{equation} 

\begin{itemize}
\item {\bf \underline{Analysis of the $\theta$ equation}}\\
In this section we simplify the angular equation of motion by defining a new variable $u=\cos\theta$. Then the angular equation 
\ref{S3-4} is given by, 
\begin{align}
\Bigg(\frac{\tilde{\rho}}{E}\Bigg)^2\dot{u}^2=\eta-u^2(\eta+\xi^2-a^2)-a^2u^4=\mathcal{G}(u(\theta))
\label{S3-7}
\end{align}
Note that the left hand side of \ref{S3-7} is positive which implies that the right hand side also needs to be positive. Since $\mathcal{G}(1)=-\xi^2$ is negative, the photon cannot access $\theta=0$. To obtain the maximum accessible value of $\theta$ denoted by $\theta_{max}$ we solve for $\mathcal{G}(u)=0$ which gives,
\begin{equation}
u^{2}=\frac{-(\eta+\xi^{2}-a^{2})\pm \sqrt{(\eta+\xi^{2}-a^{2})^{2}+4a^{2}\eta}}{2a^{2}}
\label{S3-8}
\end{equation}
If $\eta>0$ one can only consider the positive root of \ref{S3-8} since the left hand side of \ref{S3-8} is positive. Such orbits cross the equatorial plane reaching a maximum height of $\theta_{max}$ given by the solution of \ref{S3-8}. 
For negative $\eta$, we define $\eta=-|\eta|$ such that \ref{S3-8} can be rewritten as
\begin{align}
u^2=\frac{|\eta|-\xi^2+a^2\pm \sqrt{(-|\eta|+\xi^2-a^2)^2-4a^2|\eta|})}{2a^2}
\label{S3-9}
\end{align}
From \ref{S3-9} it is easy to note that for its right hand side to be positive,
\begin{align}
a^2 + |\eta|-\xi^2 >0
\label{S3-10}
\end{align}
which is the condition to be satisfied by the impact parameters. 

Finally, we note that $\eta=0$ has two solutions, namely,
\begin{eqnarray}\label{S3-11}
u_{1}^2=0 \hspace{1cm}
u_{2}^2=1-\left( \frac{\mathcal{\xi}^2}{a^2}\right)
\end{eqnarray}
If $\xi^2>a^2 $ only $u_{1}^2$ is valid else both $u_{1}^2$ and $u_{2}^2$ are valid solutions.

\item {\bf \underline{Analysis of the radial equation}}\\
In this section we consider the geodesic equation associated with the radial coordinate given by \ref{S3-3},  
\begin{align}
 \label{S3-12}
 \bigg(\frac{\tilde{\rho}}{E}\bigg)^2\dot{r}^2= a^2\xi^2 + (r(r+r_2) + a^2)^2 -4Mra\xi -\Delta (\eta + a^2 +\xi^2) = \tilde{V}(r)
 \end{align}
We will be interested in spherical photon orbits of constant radius which yields $\tilde{V}(r)=\tilde{V}'(r)=0$. Thus, we have to solve the following two equations for $\eta$ and $\xi$:
\begin{align}
a^2\xi^2 + (r(r+r_2) + a^2)^2 -4Mra\xi -\Delta (\eta + a^2 +\xi^2)=0 ~{~~\rm (obtained ~ from~ \tilde{V}(r)=0)} \nonumber \\
 2(r(r+r_{2})+a^2)(2r+r_{2})-4Ma\xi-(2r-2M+r_{2})(\eta +a^2+\xi^2)                                                                    =0 ~~~{\rm (obtained ~ from~ \tilde{V}'(r)=0)}
 \label{S3-13}
\end{align}
From \ref{S3-13} we obtain two classes of solutions for $\eta$ and $\xi$. 
\begin{enumerate}
\item 
\begin{flalign}
\eta &= -\frac{r^2(r+r_{2})^2}{a^2}\\
\xi &= a+\frac{r(r+r_{2})}{a}
\end{flalign}
\item  
\begin{flalign}
\eta&=\frac{-r^2 \left[-8a^2M(2r+r_{2})+((r+r_{2})(2r+r_{2})-2M(3r+r_{2}))^2\right]}{a^2(-2M+2r+r_{2})^2}\\
\xi&= \frac{a^2 (2(M+r)+r_{2})+r(r+r_{2})(2r+r_{2})-2M(3r+r_{2})}{a(2M-2r-r_{2})} 
\end{flalign}
\end{enumerate}
The first solution has $\eta<0$ which requires $a^2+|\eta|-\xi^2$ to be positive (see previous discussion). Substituting $\eta$ and $\xi$ from the first solution we note that $a^2+|\eta|-\xi^2=-2r(r+r_2)<0$ which makes the first solution unphysical and hence unacceptable. In the case of the second solution $\eta$ may assume any sign depending on the value of $r$ and it can be shown that the suitable conditions as discussed earlier are satisfied. We will therefore work with the second solution.

\end{itemize}

\subsection{Equation of the shadow outline}
\label{S3.3}
In this section we use the derived impact parameters from the last section to evaluate the celestial coordinates $x$ and $y$ of the black hole shadow as viewed by an observer at infinity. The position of the distant observer is taken to be $(r_0,\theta_0)$ where we take $r_0\to \infty$ and $\theta_0$ is the inclination angle of the observer. In order to obtain the outline of the black hole shadow in the observer's sky we consider the projection of the photon sphere onto the image plane.

In order to obtain the celestial coordinates we write the metric in terms of Bardeen tetrads \cite{Bardeen:1973tla,Carter:1968rr,Vries_1999}, which are associated with observers to whom the black hole appears static.

\[e^{\mu}_{(t)}=\left(\sqrt{|g^{tt}|},0,0,\frac{g^{t\phi}}{\sqrt{|g^{tt}|}} \right)\hspace{1cm}
e^{\mu}_{(r)}=\sqrt{|g^{rr}|}\left(0,1,0,0 \right) \] 
\[e^{\mu}_{(\theta)}=\sqrt{|g^{\theta\theta}|}\left(0,0,1,0 \right)\hspace{1cm}
e^{\mu}_{(\phi)}=\left(0,0,0,\sqrt{|g^{\phi\phi}|+\frac{(g^{t\phi})^2 }{|g^{tt}|}} \right) \]

From the tetrads we can compute the components of four momentum $p_{(i)}=e_{(i)}^j p_j$ of a locally inertial observer. The contravariant components of the four momentum $p^{(k)}=\eta^{(k)(l)}p_{(l)}$ of the locally inertial observer are given as,
\begin{align}
p^{(t)}=\frac{E}{c}\Bigg(c\sqrt{g^{tt}}-\xi\frac{ g^{t\phi}}{\sqrt{g^{tt}}}\Bigg) \hspace{1cm}
p^{(r)}=\pm \sqrt{\frac{\mathcal{V}(r)}{\tilde{\rho}\Delta}} \nonumber \\
p^{(\theta)}=\pm \sqrt{\frac{\Theta}{\tilde{\rho}}} \hspace{1cm}
p^{(\phi)}=\sqrt{|g^{\phi\phi}|+ \frac{(g^{t\phi})^2}{g^{tt}} }\xi \nonumber
\label{S3-15}
\end{align}

\begin{figure}[h!]%
    \centering
    \hspace{-1.5cm}
    
    \subfloat[Change in the BH shadow with dilaton parameter $r_2$. We take the inclination angle to be $\theta=45^{\circ}$ and the spin to be $a=0.6$.]{{\includegraphics[width=7.5cm]{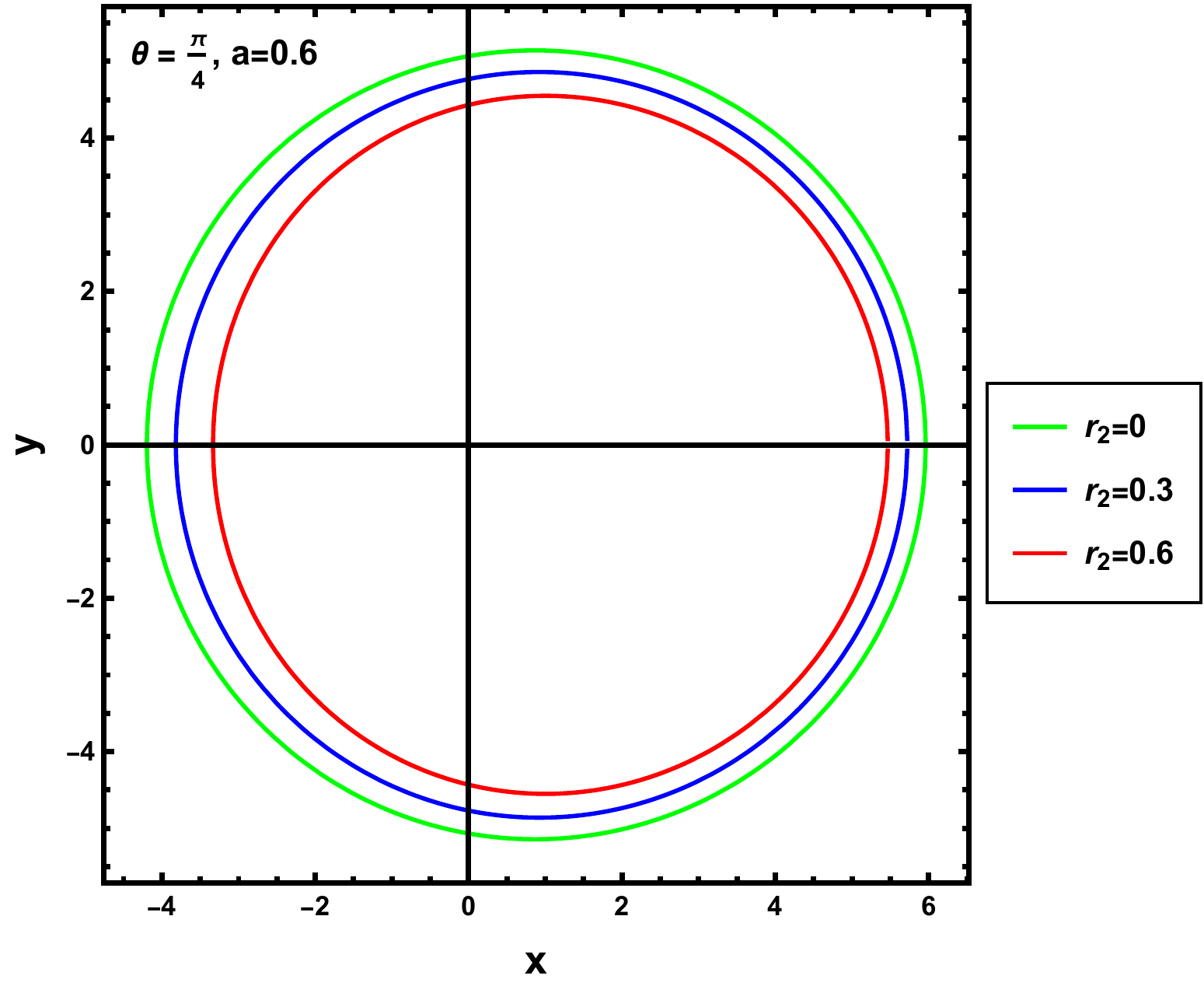} } \label{SP3_Fig_V2}}
  \qquad
     \subfloat[Change in the BH shadow with dilaton parameter $r_2$. We take the inclination angle to be $\theta=60^{\circ}$ and the spin to be $a=0.6$. ]{{\includegraphics[width=7.5cm]{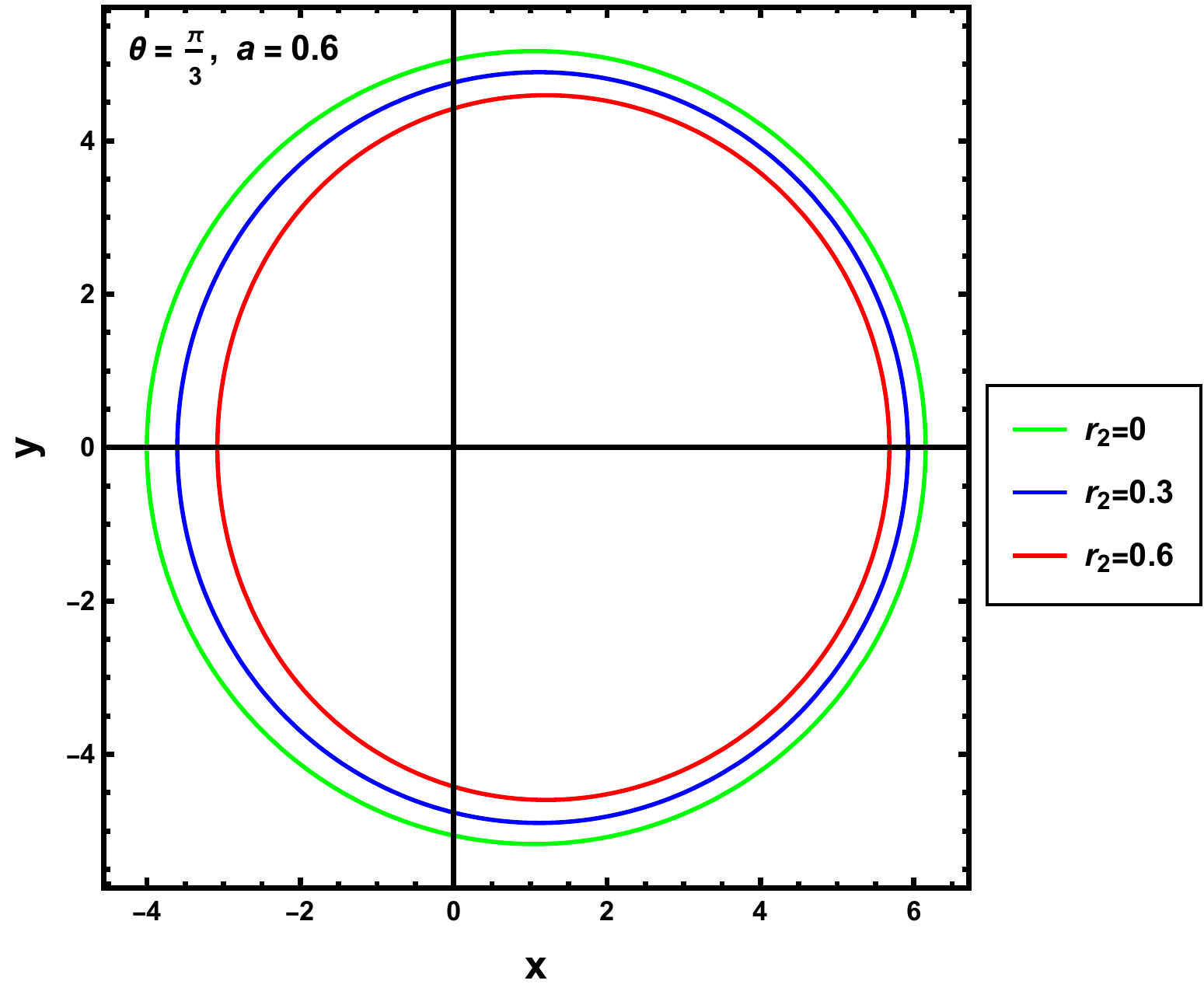} } \label{SP3_Fig_Vk}} 
   \qquad
     \subfloat[Change in the BH shadow with spin-parameter $a$. We take the inclination angle as $\theta=60^\circ$ and $r_2=0.3$ ]{{\includegraphics[width=7.5cm]{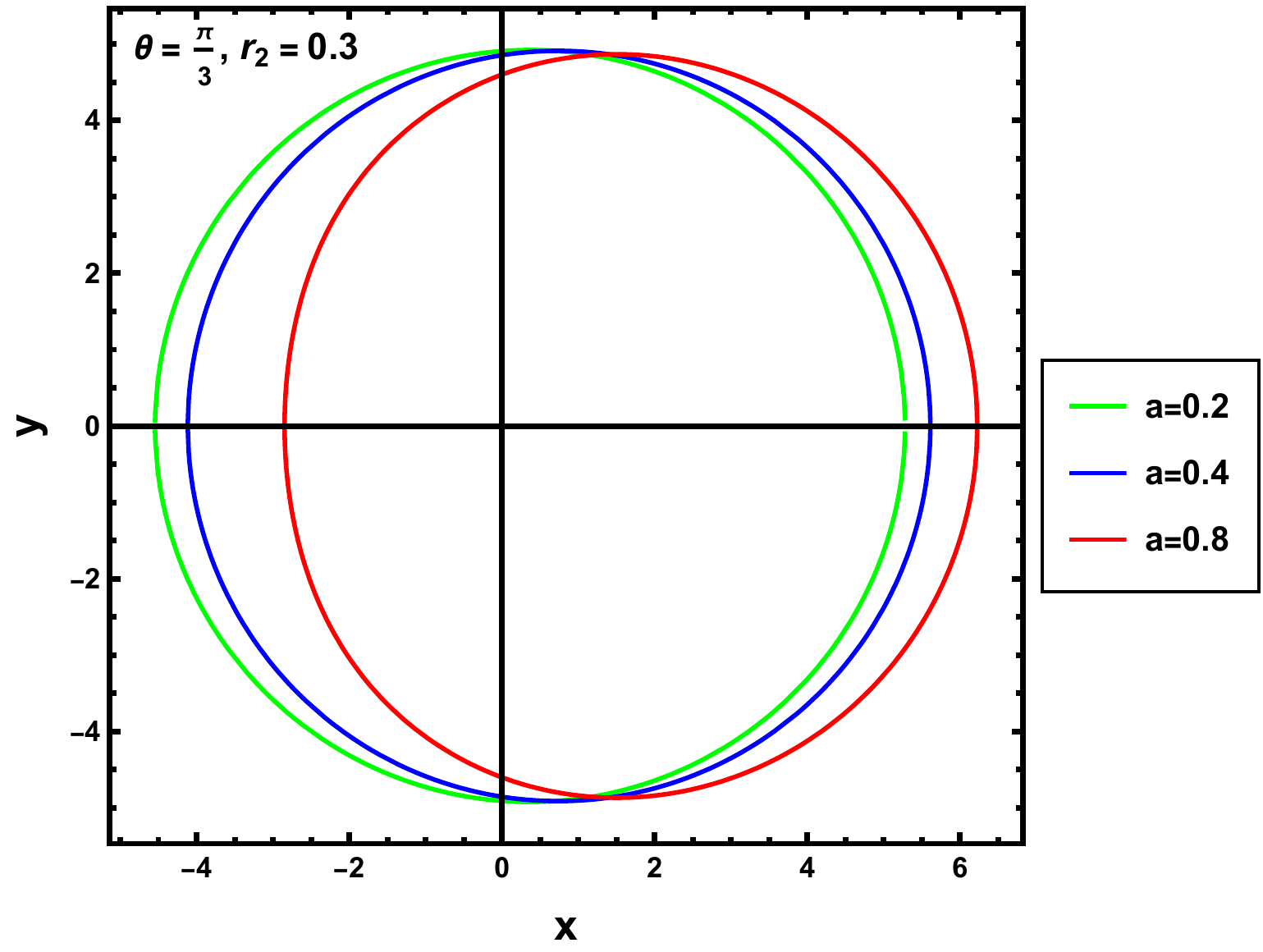} } \label{SP3_Fig_Va}}
   \qquad
    \subfloat[Change in the BH shadow with inclination angle $\theta$. Here we take the dilaton parameter $r_2=0.3$ and spin $a=0.6$.]{{\includegraphics[width=7.5cm]{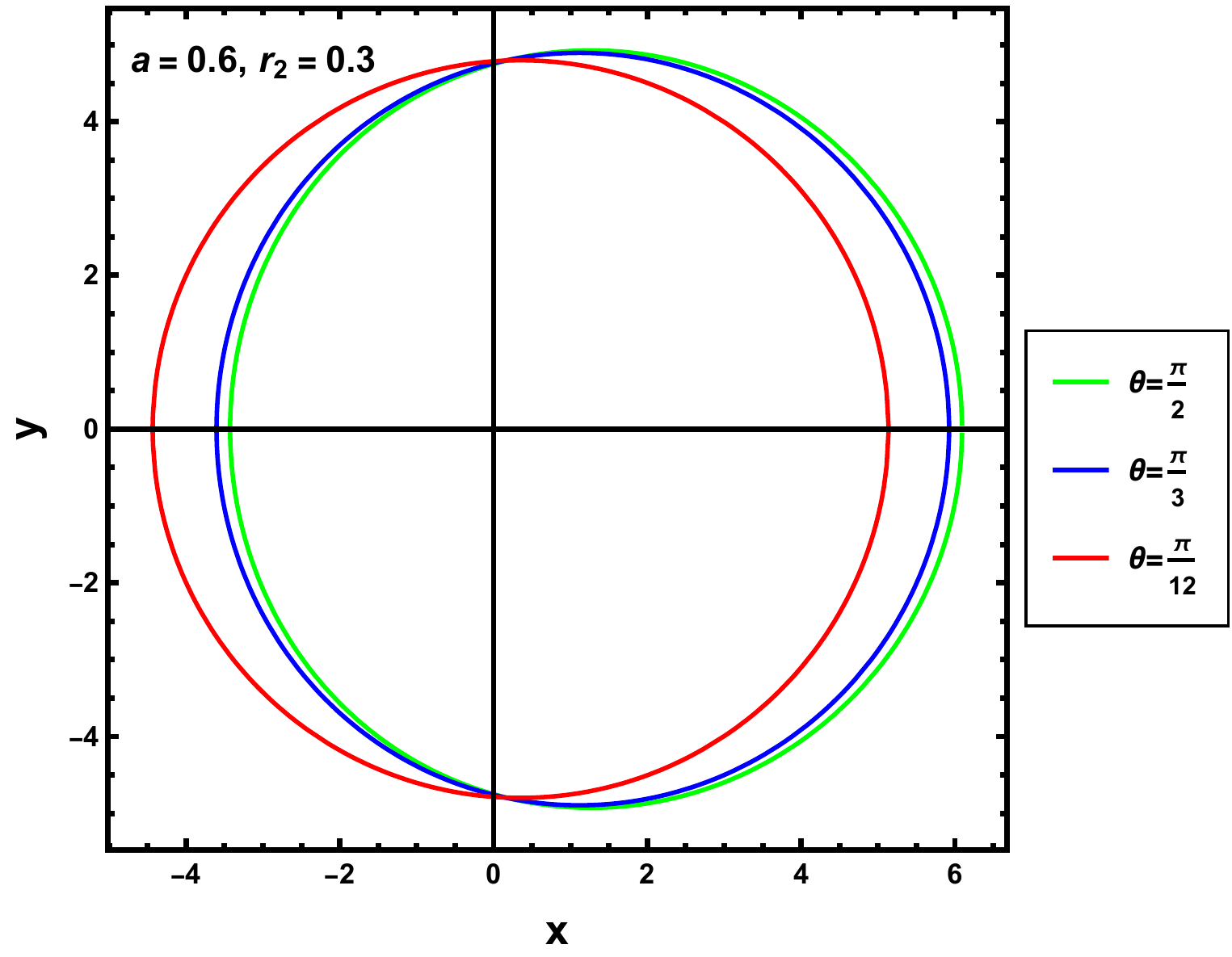} } \label{SP3_Fig_V4}}
   \caption{The above figure liiustrates the variation in the shadow structure with the dilaton parameter $r_2$, the spin parameter $a$ and the inclination angle $\theta$.
   }
    \label{SP3_Fig_V}
\end{figure}

Distant observer located at ($r_0, \theta_0$) will find the local apparent velocities of a photon to be $v_{(\theta)}=\frac{p^{(\theta)}}{p^{(r)}}$ and $v_{(\phi)}=\frac{p^{(\phi)}}{p^{(r)}}$ in which case the apparent perpendicular distance from the axis of rotation and the equatorial plane are respectively given by $d_{\phi}=r_{0}v_{(\phi)}$ and $d_{\theta}=r_{0}v_{(\theta)}$. 
These are associated with the celestial coordinates $x$ and $y$ such that 
\begin{align}
x= \lim_{r_{0}\to\infty}r_{0}v_{(\phi)} =\lim_{r_{0}\to\infty} \frac{r_{0}p^{(\phi)}(r_{0},\theta_0)}{p^{(r)}(r_{0},\theta_0)}=-\frac{\xi}{\sin \theta_0} \\
y= \lim_{r_{0}\to\infty}r_{0}v_{(\theta)}  =\lim_{r_{0}\to\infty} \frac{r_{0}p^{(\theta)}(r_{0},\theta_0)}{p^{(r)}(r_{0},\theta_0)}=\pm \sqrt{\Theta(\theta_0)}
\end{align}

\ref{SP3_Fig_V} illustrates the variation of the shape and size of the black hole shadow with the dilaton parameter $r_2$, inclination angle $\theta$, and the black hole spin $a$. The figure reveals that the shadow size decreases with an increase in the magnitude of the dilaton parameter $r_2$. We further note that when $a$ and $\theta$ are enhanced the shadow becomes increasingly non-circular \cite{Tsukamoto:2014tja,Balart:2014cga,Amir:2016cen,Tsukamoto:2017fxq}.

\section{Comparison with observations and constrains on the dilaton parameter}
\label{Sec4}
In this section we aim to constrain the Kerr-Sen parameter \(r_{2}\) using observations of M87* and SgrA* by the EHT collaboration. In order
to obtain constraints on the Kerr-Sen parameter \(r_{2}\) we theoretically calculate the observables, namely, the angular diameter \(\Delta \theta\), the axis ratio \(\Delta A\)
and the deviation from circularity \(\Delta C\)\cite{Bambi:2019tjh} for the black hole shadow, assuming the spacetime around the black hole to be described
by the Kerr-Sen metric. In our approach we use measurements for distance \(D\), mass \(M\) and the inclination angle \(\theta_{0}\) (angle between the line of sight and the jet axis) of the black hole determined from previous observations.
The observables related to black hole shadow which will be
used to find best estimate on the Kerr-Sen parameter \(r_{2}\) are discussed below:
\begin{figure}
\centering
\includegraphics[scale=0.6]{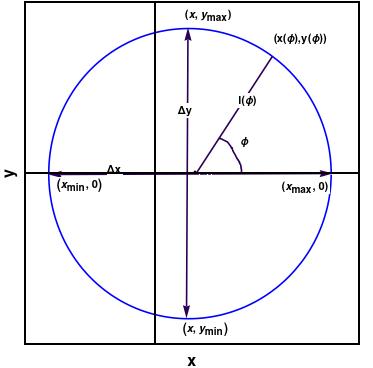}
\caption{Schematic diagram of shadow outline.}
\label{fig:sh}
\end{figure}

\paragraph{Angular diameter of shadow \(\Delta\theta\):}It is a measure of the angular width of the shadow. If the maximum width
of the shadow is \(\Delta y\) (also called the major axis length), mass of the black hole is \(M\) and distance of the black hole
from the observer is \(D\) then the angular diameter of the shadow \(\Delta \theta\)\cite{Bambi:2019tjh} is defined as:
\begin{equation}
\label{eqn:ad}
\Delta \theta=\frac{GM\Delta y}{c^{2}D}
\end{equation}
The value of \(\Delta y\) is calculated from the equation of the shadow which contains the impact parameters
\(\xi\) and \(\eta\). The impact parameters in turn depend on the metric components $r_2$, $a$ and the inclination angle $\theta_0$.  
Therefore, the
angular diameter also depends on the three aforesaid parameters and thus \(r_{2}\) can be constrained
using experimental observations of $\Delta \theta$ for predetermined inclination angle \(\theta_{0}\).
\paragraph{Axis ratio \(\Delta A\) of the black hole shadow:}As the shadow of the black hole is in general not circular the major
axis \(\Delta y\) and the minor axis \(\Delta x\) may not be equal. From \ref{fig:sh}, the axis ratio \(\Delta A\)
is defined as\cite{Bambi:2019tjh}:
\begin{equation}
\label{eqn:ar}
\Delta A=\frac{\Delta y}{\Delta x}
\end{equation}
where the minor axis \(\Delta x\) is also calculated from the equation of the shadow and hence, \(\Delta A\) also depends on \(r_{2}\), \(a\) and \(\theta_{0}\).
\paragraph{Deviation from Circularity \(\Delta C\):}Deviation from circularity \(\Delta C\) measures the amount of deviation from the circular
shape of the shadow\cite{Bambi:2019tjh}. It is  defined as follows:
\begin{gather}
\label{eqn:dc}
\Delta C=\frac{1}{R_{avg}}\sqrt{\frac{1}{2\pi}\int^{2\pi}_{0}\{l(\phi)-R_{avg}\}^{2}d\phi}\\
\text{Here,}\ R_{avg}=\sqrt{\frac{1}{2\pi}\int^{2\pi}_{0}l(\phi)^{2}d\phi}\\
l(\phi)=\sqrt{(x(\phi)-x_{c})^{2}+y(\phi)^{2}}
\end{gather}
In the above expression \(R_{avg}\) is the average radius of the shadow. \(l(\phi)\) is the length of the line joining the point
\((x(\phi), y(\phi))\) on the shadow and the geometric centre \((x_{c}, 0)\) (see \ref{fig:sh}). It must be
noted that due to reflection symmetry of the Kerr-Sen metric, the shape of the shadow is symmetric about the \(x\)-axis, hence the \(y\) coordinate of the geometric center is
0. The \(x\) coordinate of the geometric centre is calculated using the formula,
\begin{equation}
x_{c}=\frac{\int^{2\pi}_{0}x(\phi)dS}{\int^{2\pi}_{0}dS}\ ( \text{here \(dS\) is the area element})
\end{equation}


\paragraph{EHT observations of M87*:} The EHT collaboration measured the angular diameter \(\Delta \theta\), the axis ratio \(\Delta A\) and the deviation from circularity \(\Delta C\) for the image of M87*, the supermassive black hole candidate at the center of the galaxy M87 \cite{EventHorizonTelescope:2019dse,EventHorizonTelescope:2019ggy,EventHorizonTelescope:2019pgp}. The values reported are given below:
\begin{enumerate}
\item \(\Delta\theta\) = \((42 \pm 3)\mu as\). The EHT also reports a maximum offset of 10\% between the shadow angular diameter and the image angular diameter. Thus, the shadow angular diameter can be as small as \(\Delta\theta = (37.8 \pm 2.7)\mu as\)\cite{EventHorizonTelescope:2019dse,EventHorizonTelescope:2019ggy,EventHorizonTelescope:2019pgp}.
\item \(\Delta A \lesssim 4/3\)\cite{EventHorizonTelescope:2019dse,EventHorizonTelescope:2019ggy,EventHorizonTelescope:2019pgp}.
\item \(\Delta C \lesssim 10\% \)\cite{EventHorizonTelescope:2019dse,EventHorizonTelescope:2019ggy,EventHorizonTelescope:2019pgp}.
\end{enumerate}
In order to determine the observationally favored Kerr-Sen parameter \(r_{2}\), we need to theoretically derive the above three observables as functions of the metric parameters $r_2$ and $a$. As evident from \ref{eqn:ad} a theoretical computation of the angular diameter $\Delta \theta$ requires independent measurements of the black hole mass, distance and inclination (required to derive $\Delta y$). We use previously estimated masses and distance of this source to compute the theoretical angular diameter. 
The distance of M87* as reported from stellar population measurements turns out to be \(D = (16.8 \pm 0.8)\) Mpc\cite{Bird:2010rd,Blakeslee:2009tc,Cantiello:2018ffy}. The angle of inclination which is the angle between the line of sight and the jet axis (the jet axis is believed to coincide with the spin axis of the black hole) is \(17^{\circ}\)\cite{CraigWalker:2018vam}. The mass of M87* has been measured using different methods. The mass measurement by modelling surface brightness and dispersion in stellar velocity was found to be \(M=6.2^{+1.1}_{-0.6}\times 10^{9} M_{\odot}\)\cite{EventHorizonTelescope:2019ggy,Gebhardt:2009cr,Gebhardt:2000fk}. Mass measurements from kinematic study of gas disk gives \(M=3.5^{+0.9}_{-0.3}\times10^{9} M_{\odot}\)\cite{EventHorizonTelescope:2019ggy,Walsh:2013uua}. Mass measured from the image of M87* by the EHT collaboration assuming \gr\ turns out to be \(M =
(6.5 \pm 0.7)\times 10^{9}M_{\odot}\)\cite{EventHorizonTelescope:2019dse,EventHorizonTelescope:2019ggy,EventHorizonTelescope:2019pgp}. 

\paragraph{EHT observations of SgrA*:} In May 2022 the EHT collaboration released the image of the black hole SgrA* present at the galactic center of the Milky Way galaxy. The angular diameter of the image is found to be \(\Delta \theta= (51.8 \pm 2.3)\mu as\)\cite{EventHorizonTelescope:2022wkp,EventHorizonTelescope:2022apq,EventHorizonTelescope:2022wok,EventHorizonTelescope:2022exc,EventHorizonTelescope:2022urf,EventHorizonTelescope:2022xqj}. The angular diameter of the shadow is \(\Delta\theta = (48.7 \pm 7)\mu as\)\cite{EventHorizonTelescope:2022wkp,EventHorizonTelescope:2022apq,EventHorizonTelescope:2022wok,EventHorizonTelescope:2022exc,EventHorizonTelescope:2022urf,EventHorizonTelescope:2022xqj}.\\
The mass and distance of SgrA* reported by the Keck collaboration keeping the redshift parameter free, are \(M = (3.975\pm
0.058 \pm 0.026) \times 10^{6}M_{\odot}\)\cite{Do:2019txf} and \(D = (7959 \pm 59 \pm 32)\)pc\cite{Do:2019txf} respectively. Fixing the value of redshift parameter to unity the
mass and distance reported by the Keck team are \(M = (3.951 \pm 0.047) \times 10^{6}M_{\odot}\) and \(D = (7935 \pm 50)\)pc. The mass and distance of Sgr A* reported by the GRAVITY collaboration are \(M = (4.261 \pm 0.012) \times 10^{6}M_{\odot}\) and \(D =
(8246.7 \pm 9.3)\)pc\cite{GRAVITY:2021xju,GRAVITY:2020gka} respectively. When systematics due to optical aberrations are taken into account the GRAVITY collaboration constrains the mass and distance of Sgr A* to $M=4.297 \pm 0.012 \pm 0.040\times 10^6 M_\odot$ and $D=8277 \pm 9 \pm 33$ pc respectively. Apart from mass and distance we also need to provide independent measurements of the inclination angle to establish observational constrains on $r_2$. From \cite{refId0} we take $\theta \simeq 134^\circ$ (or equivalently $46^\circ$). When models based on extensive numerical simulations are compared with the the observed image of Sgr A*, one concludes that the inclination angle of the source is $\theta<50^\circ$.
The estimates for axes ratio \(\Delta A\) and the deviation from circularity \(\Delta C\) for image of SgrA* by EHT collaboration are yet to be released, hence, for SgrA* the observable \(\Delta \theta\) will only  be used for estimating \(r_{2}\).\\

To constrain the Kerr-Sen/dilaton parameter $r_2$ using EHT observations, we proceed with the following approach:
\begin{enumerate}
\item We derive the outline of the black hole shadow by calculating the impact parameters \(\xi\) and \(\eta\) for the Kerr-Sen
black hole. We obtain the parametric equations of the black hole shadow outline.
\item We fix the value of \(r_{2}\) and vary the spin \(a\) of black hole in suitable range such that event horizon radius is real and positive.
\item For each combination of \((r_{2}, a)\) we calculate the values of angular diameter \(\Delta \theta\), axis ratio \(\Delta A\) and the deviation from circularity \(\Delta C\). In these calculations we use values of mass \(M\), distance \(D\) and inclination angle \(\theta_{0}\) from previous measurements as discussed above.
\item Then we repeat steps 2 and 3 for different values of \(r_{2}\) upto \(r_{2} = 1.8\). It must be noted that the dilaton parameter \(r_{2}\) varies in the range \(0 \leq r_{2} \leq 2\) since the horizon radius (in units of $M$) is given by $r_h={1}-\frac{r_2}{2} +\sqrt{\bigg({1}-\frac{r_2}{2}\bigg)^2 - a^2}$ which needs to be real and positive.
\item After obtaining values for $\Delta\theta$, $\Delta A$ and $\Delta C$ we plot contour plots for angular diameter \(\Delta \theta\), density
plots for axes ratio \(\Delta A\) and deviation from circularity \(\Delta C\) as functions of \(r_{2}\) and $a$.
\item The values of \(r_{2}\) which are able to reproduce the observationally measured \(\Delta\theta, \Delta A\) and \(\Delta C\) give us the estimates of \(r_{2}\) based on shadow related measurements. 
\end{enumerate}

\paragraph{Constrains on the dilaton parameter \(r_{2}\) from EHT observations of M87* :} 
Here we discuss the observationally favored magnitude of the dilaton parameter $r_2$ derived from the shadow of M87* released by the Event Horizon Telescope collaboration in April 2019. 
In order to get an understanding of the observationally preferred value of $r_2$ we theoretically compute the observables, namely, $\Delta \theta$ (angular diameter), $\Delta A$ (axis ratio) and $\Delta C$ (deviation from circularity) related to the black hole shadow, which have been discussed towards the beginning of this section. It is important to recall that these observables depend on the metric parameters $r_2$, $a$ and the inclination angle $\theta_0$. In addition, the theoretically derived $\Delta \theta$ requires independent estimates of the mass $M$ and the distance $D$ of the black hole (see \ref{eqn:ad}). The inclination angle is assumed to be $17^\circ$ and the distance $D$ is taken to be $D=16.8$ Mpc (obtained from stellar population measurements) throughout this discussion.

\begin{figure}[t!]
\begin{center}
\qquad
\subfloat[\label{Fig3a} Contours showing variation of the angular diameter of the shadow of M87* with the dilaton parameter $r_2$ and the spin parameter $a$. In order to compute the contours the mass $M\simeq 3.5\times 10^9 M_\odot$ and distance $D\simeq16.8~\textrm{Mpc}$
is considered.]{\includegraphics[scale=0.295]{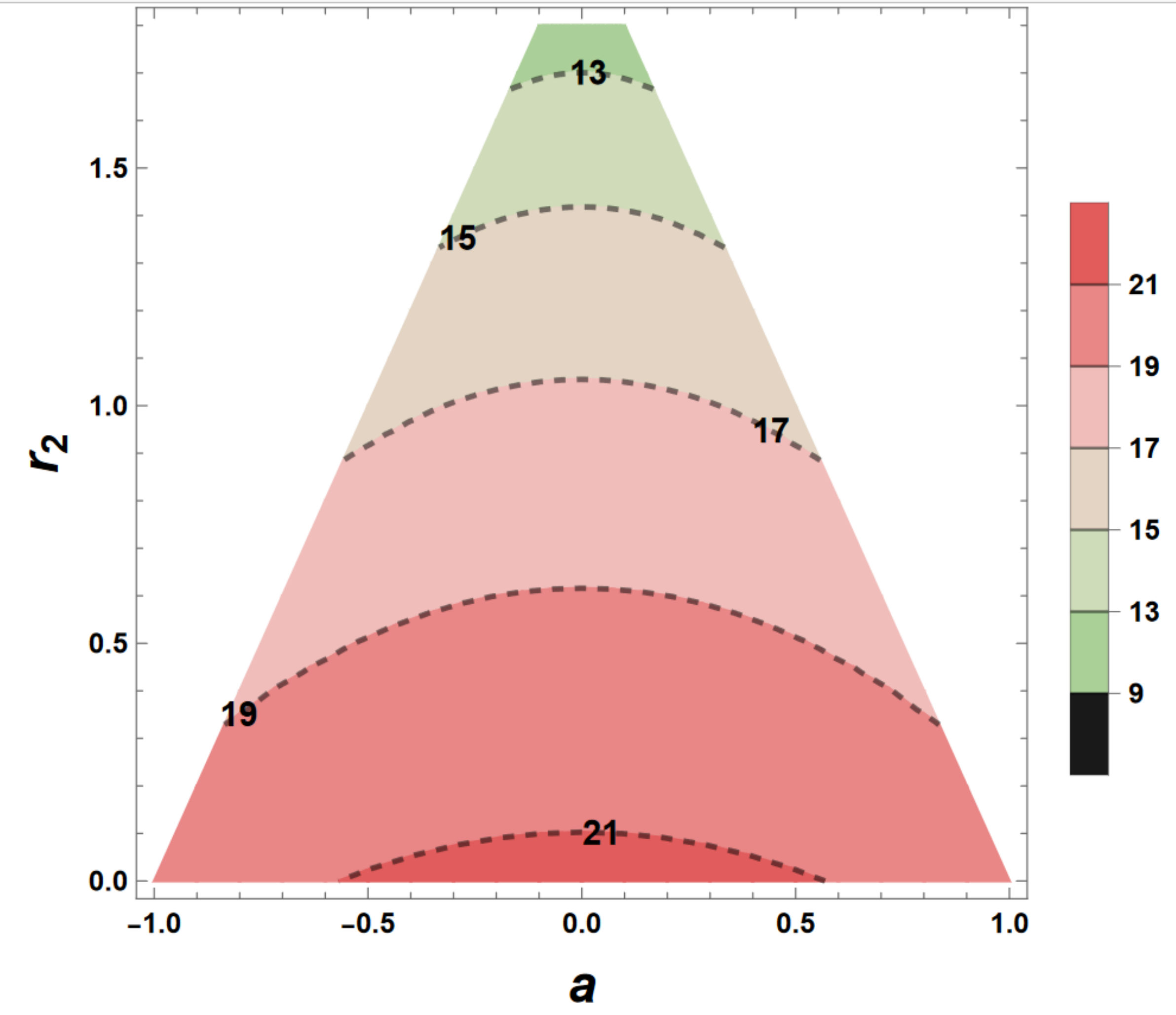}}
\qquad
\hspace{0.2cm}
\subfloat[Contours showing variation of the angular diameter of the shadow of M87* with the dilaton parameter $r_2$ and the spin parameter $a$. In order to compute the contours the mass $M\simeq 6.2\times 10^9 M_\odot$ and distance $D\simeq16.8~\textrm{Mpc}$ is considered.
]{\includegraphics[scale=0.295]{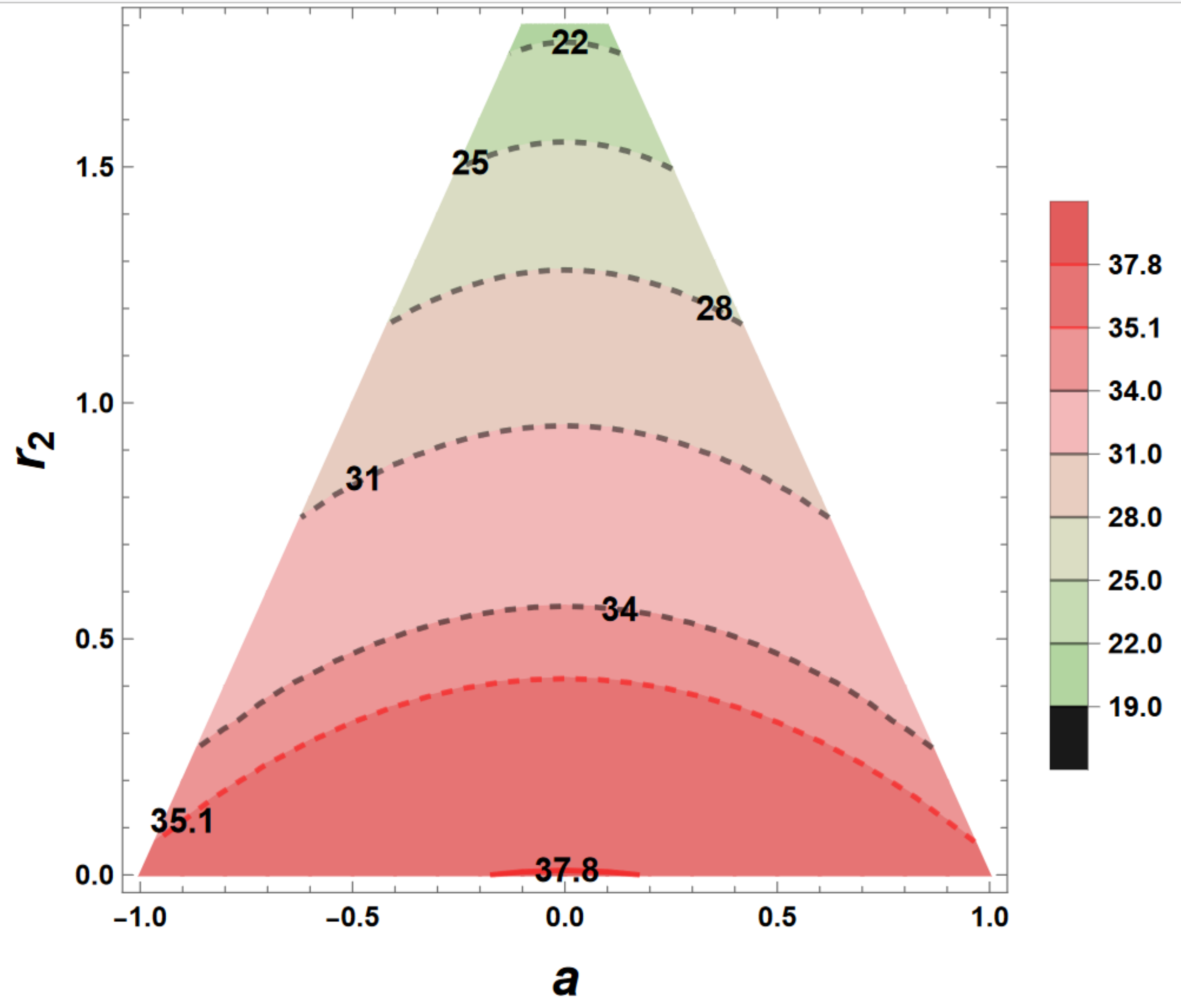}\label{Fig3b}
}

\caption{The above figure illustrates the dependence of the angular diameter of the shadow of M87* on the dilaton parameter $r_2$ and the spin parameter $a$ assuming previously estimated distance and mass.}
\label{Fig3}
\end{center}
\end{figure}

In \ref{Fig3a} we plot the theoretical angular diameter $\Delta \theta$ of M87* as functions of $r_2$ and $a$ assuming mass $M\simeq 3.5\times 10^9 M_\odot$ (obtained from gas dynamics studies).
We note from \ref{Fig3a} that there is no suitable $r_2$ in the range 0 to 2 (obtained from the considerations of a real, positive event horizon) which can reproduce the observed angular diameter of M87* denoted by $\Phi= 42\pm 3 \mu as$. Even when the maximum offset of 10\% in the angular diameter is considered i.e $\Phi=37.8 \pm 2.7 \mu as$ is taken as the observed value, the mass $M\simeq 3.5\times 10^9 M_\odot$ falls short in addressing the observations. Since the theoretical angular diameter is directly proportional to the mass (\ref{eqn:ad}) therefore a larger mass of the source is required to reproduce the observations. Hence, it seems that the mass of M87* measured from gas dynamics studies needs to be revisited. 

We next consider calculating the theoretical angular diameter of shadow of M87* keeping the distance fixed to  $D=16.8$ Mpc but using the mass $M\simeq 6.2\times 10^9 M_\odot$ obtained from stellar dynamics measurements. With these values of mass and distance we evaluate the theoretical angular diameter $\Delta \theta$ for M87* which is plotted in \ref{Fig3b}. 
From the figure it is evident that no value of $r_2$ can reproduce the observed image diameter $\Phi= 42\pm 3 \mu as$. However, when maximum offset of 10\% in the angular diameter is considered, $0.1\lesssim r_2 \lesssim 0.3$ is required to explain the observed angular diameter within 1-$\sigma$ ($35.1\mu as=(37.8 -2.7)\mu as$, denoted by the red dashed line in \ref{Fig3b}). Thus, when angular diameter is calculated with mass $M\simeq 6.2\times 10^9 M_\odot$ a non-zero $r_2$ can only explain the observations within 1-$\sigma$ if maximum offset of 10\% in the image diameter is allowed.

\begin{figure}[t!]
\begin{center}
\includegraphics[scale=0.295]{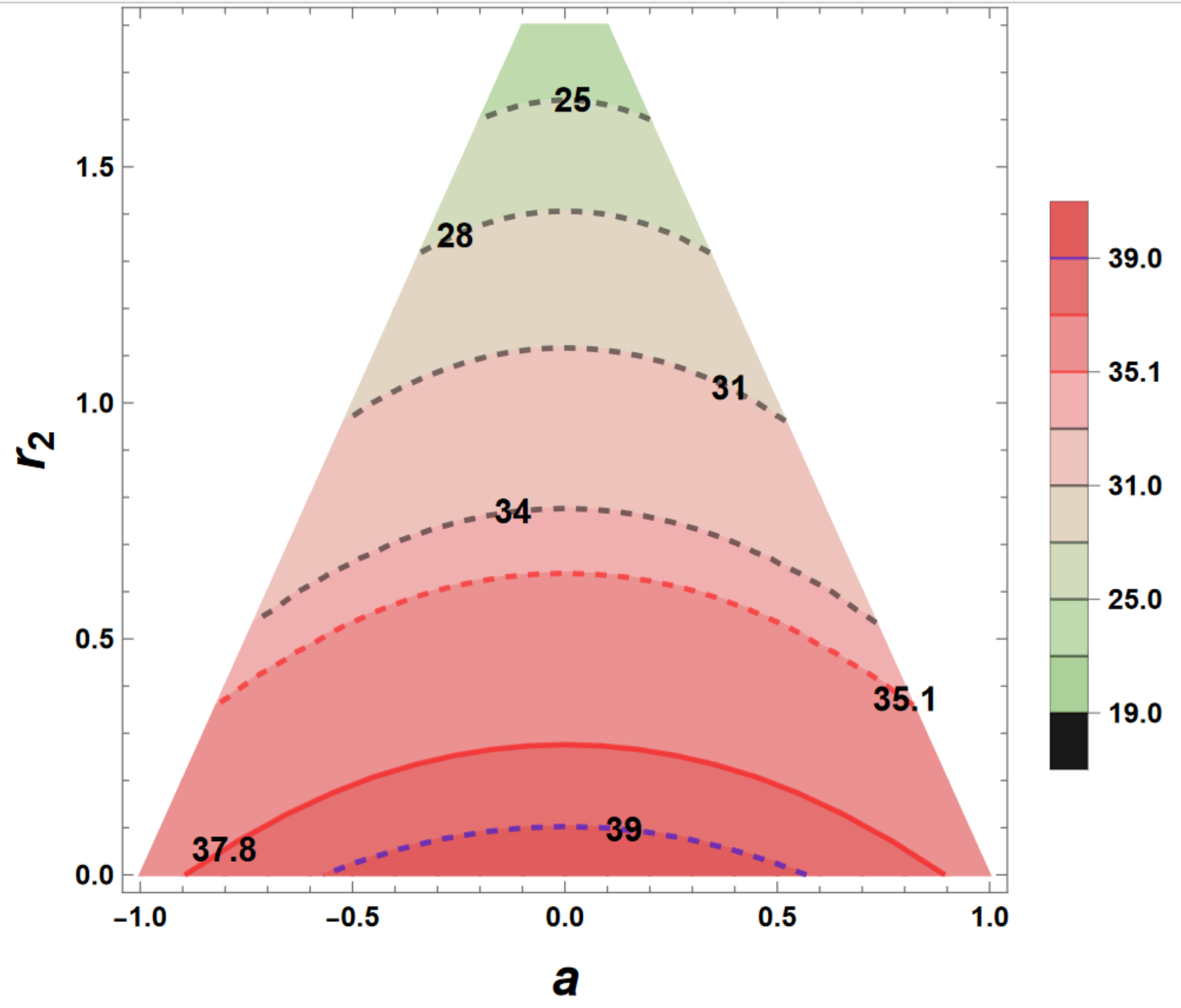}
\caption{The above figure demonstrates the dependence of the angular diameter of the shadow of M87* on the dilaton parameter $r_2$ and the spin parameter $a$ assuming distance $D\simeq16.8\textrm{Mpc}$ and mass $M\simeq 6.5\times 10^9 M_\odot$. This mass is derived by the EHT team from the shadow diameter assuming GR. Therefore, we cannot use this mass to constrain parameters of another alternate gravity theory. The contours with $M\simeq 6.5\times 10^9 M_\odot$ are plotted for purpose of comparison and completeness only. 
}
\label{Fig4}
\end{center}
\end{figure}

For completeness we also calculate the theoretical angular diameter with mass $M\simeq 6.5\times 10^9 M_\odot$ which is the mass derived by the EHT collaboration from the observed shadow of M87* assuming general relativity. Since this is the largest among all the three masses, it can explain the the observed image diameter of $\Phi= 42\pm 3 \mu as$ within 1-$\sigma$ (39 $\mu as$ denoted by the blue dashed line in \ref{Fig4}). If the maximum offset of 10\% is allowed then a non-zero dilaton charge $0\lesssim r_2 \lesssim 0.2$ can explain the observed shadow diameter ($\Phi=37.8\mu as$, denoted by the red solid line in \ref{Fig4}). 
However, $M\simeq 6.5\times 10^9 M_\odot$ should not be used to infer the observationally favored magnitude of $r_2$ since this mass is obtained from shadow measurements assuming GR. Therefore, using this mass estimate we cannot constrain another alternative gravity theory.

\begin{figure}[t!]
\begin{center}
\qquad
\subfloat[\label{Fig5a} Figure illustrating the dependence of the deviation from circularity $\Delta C$ on $r_2$ and $a$. Here $\Delta C$ has been calculated assuming the inclination angle $\theta_0=17^\circ$ corresponding to M87*.]{\includegraphics[scale=0.261]{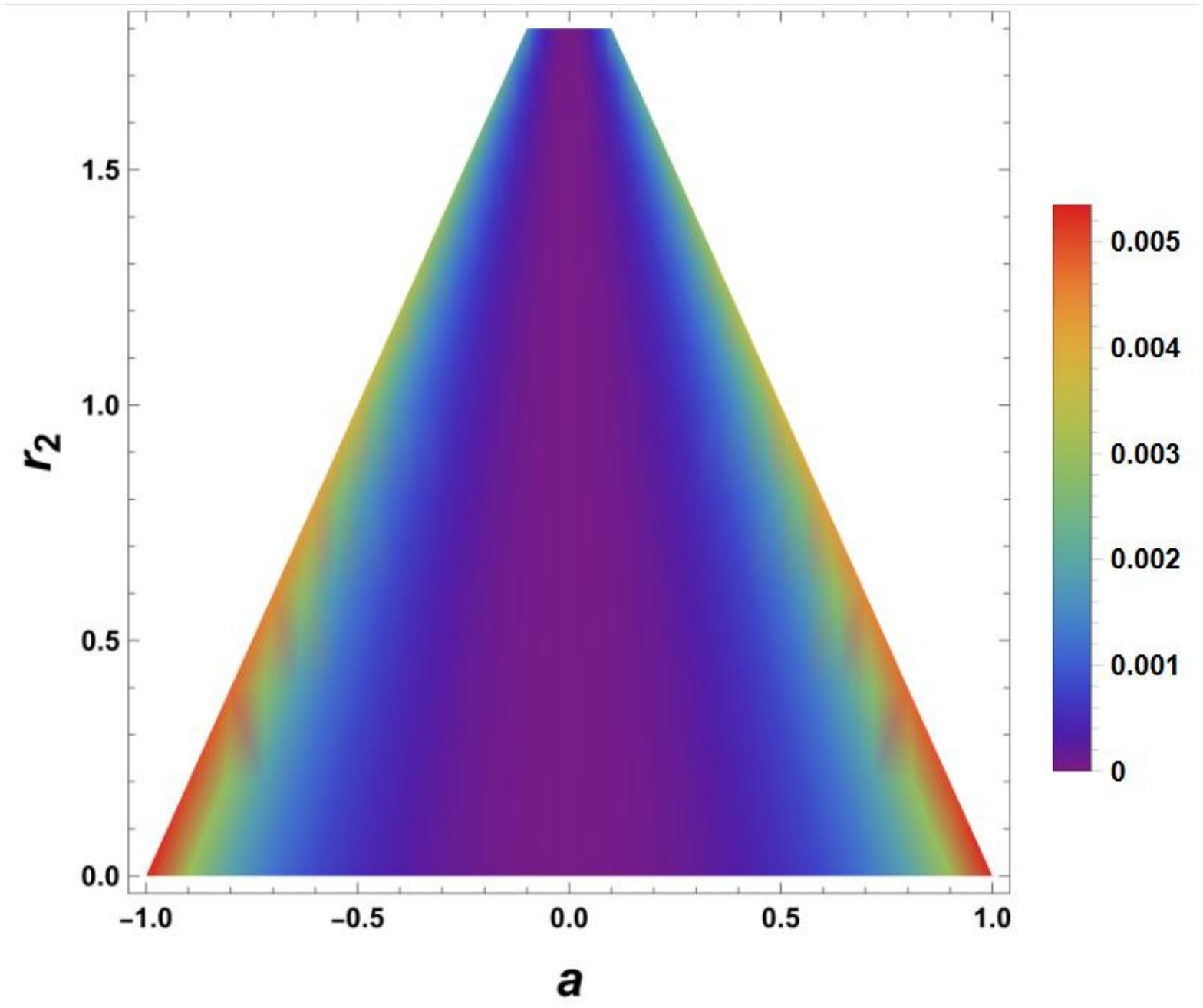}}
\qquad
\hspace{0.2cm}
\subfloat[Figure illustrating the dependence of the axis ratio $\Delta A$ on $r_2$ and $a$. Here $\Delta A$ has been calculated assuming the inclination angle $\theta_0=17^\circ$ corresponding to M87*.]{\includegraphics[scale=0.258]{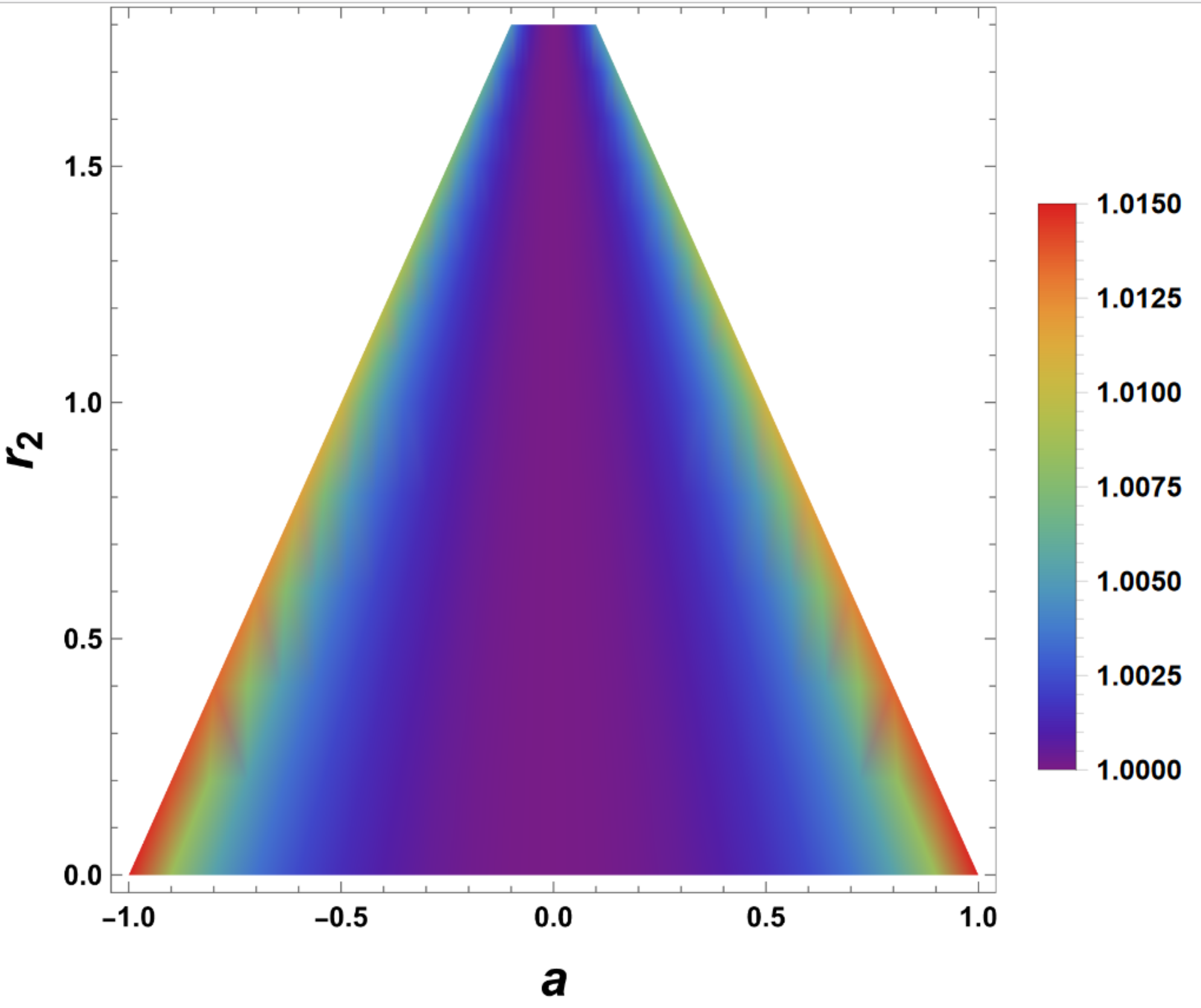}\label{Fig5b}
}

\caption{The above figure depicts the variation of $\Delta C$ and $\Delta A$ for M87* with the dilaton parameter $r_2$ and the spin parameter $a$. }
\label{Fig5}
\end{center}
\end{figure}

We now discuss the constrains on $r_2$ from the other two observables $\Delta C$ and $\Delta A$. The theoretical computation of these two observables does not require information about the mass and distance of the source. One however needs to provide the inclination angle of the source (which in the present case in $17^\circ$) to obtain $\Delta C$ and $\Delta A$ as functions of $r_2$ and $a$. According to the EHT results, the deviation from circularity \(\Delta C\lesssim10\%\)\cite{EventHorizonTelescope:2019dse,EventHorizonTelescope:2019ggy,EventHorizonTelescope:2019pgp} or 0.1 for M87*. The density plot of \(\Delta C\) for M87* is shown in \ref{Fig5a}.
From the density plot we observe that for all values of spin and \(r_{2}\) \(\Delta C<10\%\) or 0.1 is realized. Thus, \(\Delta C\) estimate for M87* does not give any additional bound on the Kerr-Sen parameter \(r_{2}\).
The EHT collaboration estimates an upper bound on the axis ratio \(\Delta A\) for the image of M87*, i.e, \(\Delta A  <\frac{4}{3}\)\cite{EventHorizonTelescope:2019dse,EventHorizonTelescope:2019ggy,EventHorizonTelescope:2019pgp}. The density plot for axes ratio \(\Delta A\) in \ref{Fig5b} indicate that for all values of \(r_{2}\) and \(a\) the axes ratio \(\Delta A<\frac{4}{3}\). Thus, EHT estimate of axes ratio \(\Delta A\) for M87* does not provide any additional constrain on the dilaton parameter \(r_{2}\). It can be said that the axis ratio estimate allows non-zero values of Kerr-Sen parameter, although it does not constrain it.\\

\paragraph{Constrains on the dilaton parameter \(r_{2}\) from EHT observations of Sgr A*:}
\begin{figure}[]
\begin{center}

\subfloat[\label{Fig6a}The above figure demonstrates the variation of the angular diameter with $r_2$ and $a$ assuming $M=3.951\times 10^6 M_\odot$ and $D=7.935$ kpc.]{\includegraphics[scale=0.267]{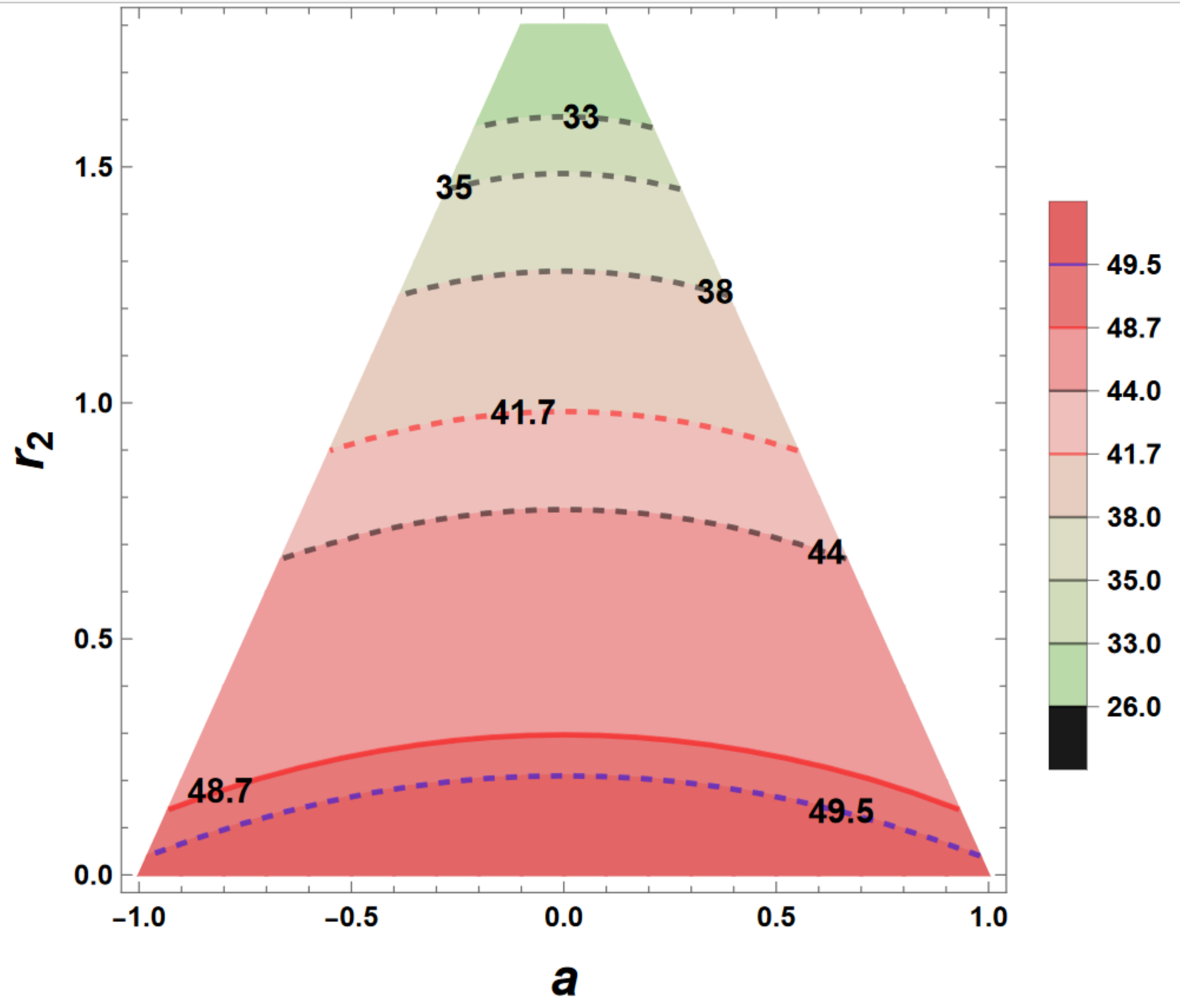}}
\hspace{0.36cm}
\subfloat[\label{Fig6b}The above figure demonstrates the variation of the angular diameter with $r_2$ and $a$ assuming $M=3.975\times 10^6 M_\odot$ and $D=7.959$ kpc.]
{\includegraphics[scale=0.267]{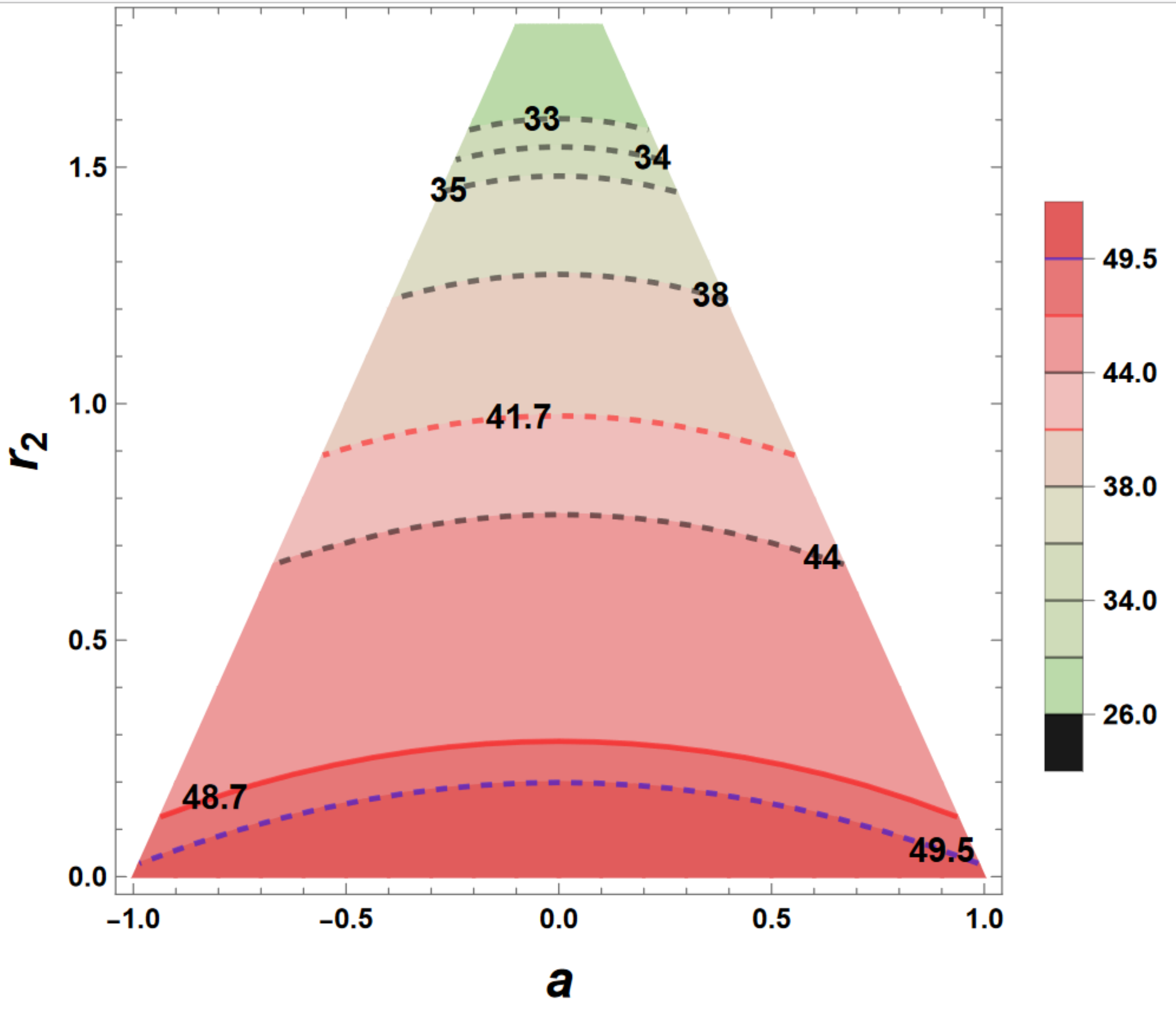}}
\\
\subfloat[\label{Fig6c}The above figure demonstrates the variation of the angular diameter with $r_2$ and $a$ assuming $M=4.261 \times 10^6 M_\odot$ and $D=8.2467$ kpc.]
{\includegraphics[scale=0.267]{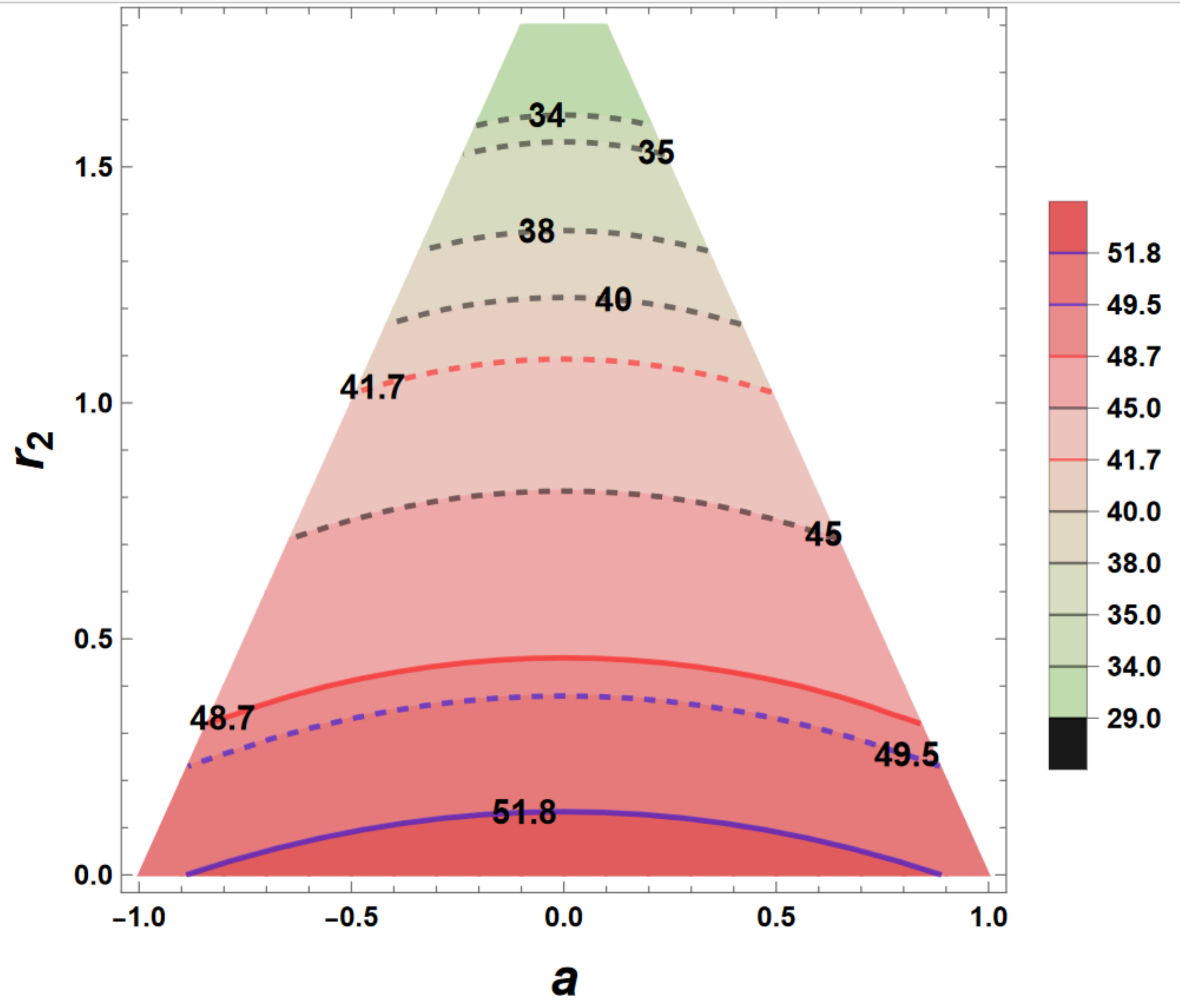}}
\hspace{0.36cm}
\subfloat[\label{Fig6d}The above figure demonstrates the variation of the angular diameter with $r_2$ and $a$ assuming $M=4.297\times 10^6 M_\odot$ and $D=8.277$ kpc.]
{\includegraphics[scale=0.267]{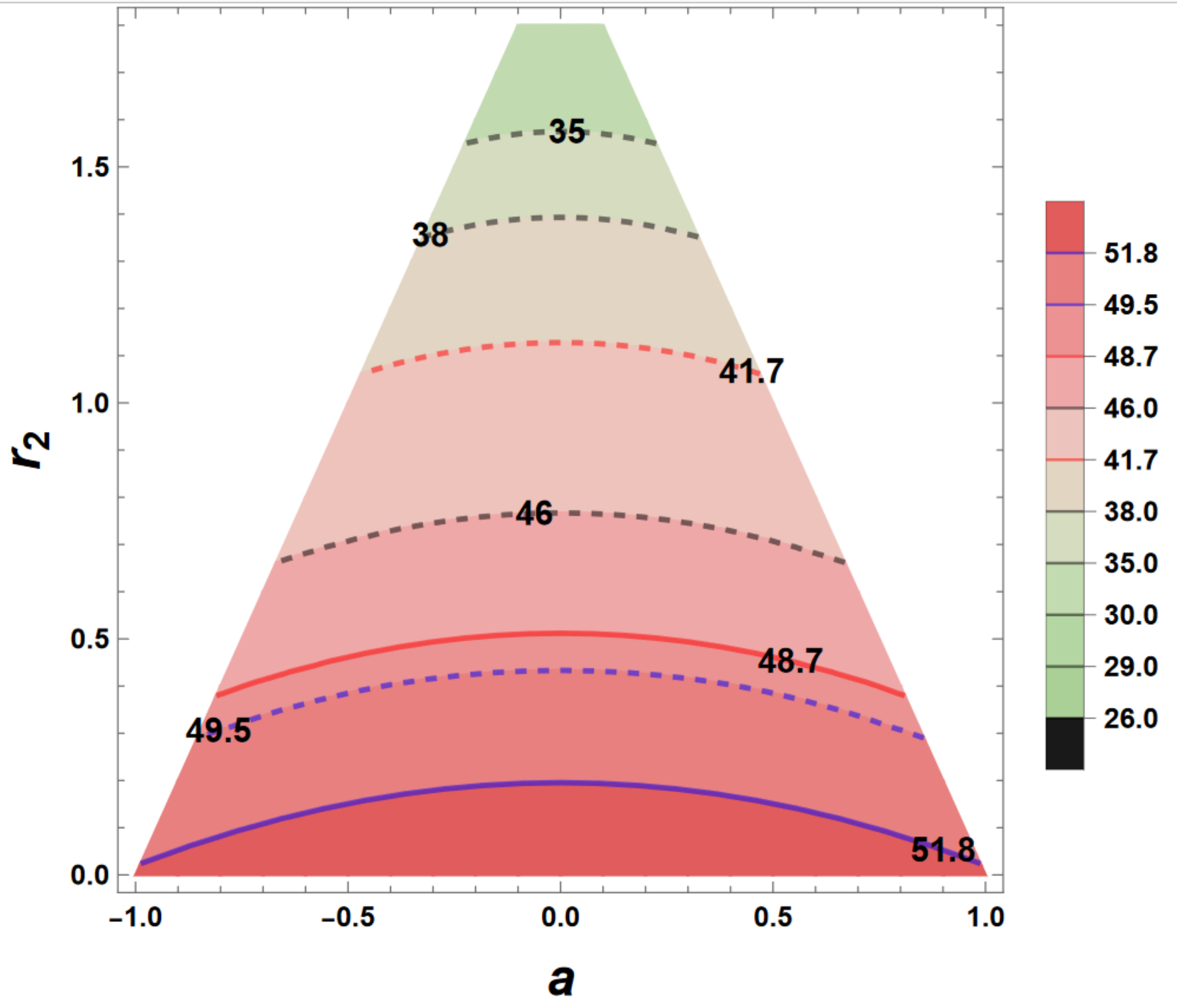}}

\caption{The figure illustrates the variation of the angular diameter of Sgr A* with metric parameters $r_2$ and $a$ assuming masses and distances reported by the Keck team and the GRAVITY collaboration. In order to compute the angular diameter the inclination angle is taken to be $\theta_0=46^\circ$.}\label{Fig6}
\end{center}
\end{figure}

The EHT collaboration measured the angular diameter for image of SgrA* to be \(\Delta \theta=(51.8\pm2.3)\mu as\) while the shadow diameter is estimated to be \(\Delta\theta=(48.7\pm7)\mu as\) \cite{EventHorizonTelescope:2022wkp,EventHorizonTelescope:2022apq,EventHorizonTelescope:2022wok,EventHorizonTelescope:2022exc,EventHorizonTelescope:2022urf,EventHorizonTelescope:2022xqj}. The theoretical angular diameter depends on the mass $M$, the distance $D$, the inclination angle $\theta_0$ and the metric parameters $r_2$ and $a$ (see \ref{eqn:ad}). As before, we use previously determined masses and distances of the source to compute the theoretical angular diameter which is then compared with the observations to establish constrains on $r_2$. The angle of inclination has an estimated upper bound \(\theta_{0}<50^{\circ}\) obtained by comparing the image of Sgr A* with extensive numerical simulations \cite{EventHorizonTelescope:2022wkp}. Following \cite{refId0} we fix the inclination angle to be \(\theta_{0}=46^{\circ}\) for the present work. 

In \ref{Fig6} the contours of theoretical angular diameter \(\Delta \theta\) of the shadow of Sgr A* are plotted for different estimates of mass and distance. The mass and distance of the source have been well constrained by the Keck team and the GRAVITY collaboration. We first discuss the constrains on $r_2$ assuming distance and mass measurements by the Keck team \cite{Do:2019txf}. Keeping the red-shift parameter free the mass and distance of Sgr A* turns out to be $M = (3.975 \pm 0.058 \pm 0.026) \times 10^6 M_\odot$ and $D = (7959 \pm 59 \pm 32) $ pc, respectively. When the red-shift parameter is fixed to unity the distance and mass estimates by the Keck team yield $D = (7935 \pm 50)$ pc and $M = (3.951 \pm 0.047)\times 10^6 M_\odot$. In \ref{Fig6a} and \ref{Fig6b} we plot the contours of angular diameter of the shadow using the masses and distances estimated by the Keck team. From the figures it is evident that the observed shadow diameter of \(\Phi=48.7\mu as\) can be reproduced by $0.1\lesssim r_2 \lesssim 0.2$ (red solid line in \ref{Fig6a} and \ref{Fig6b}). When the lower 1-$\sigma$ interval is considered, i.e., \(\Phi=(48.7-7=41.7)\mu as\), then, $r_2$ can be as high as unity (red dashed line in \ref{Fig6a} and \ref{Fig6b}). Since the error bar associated with the shadow diameter is quite high ($\pm 7 \mu as$) we do not assign much importance to this result but emphasize that a small but non-trivial value of $r_2\simeq 0.1-0.2$ is required to reproduce the central value of the observed shadow diameter. We further note that $0\lesssim r_2 \lesssim 0.1$ can explain the observed image diameter within 1-$\sigma$, (\(\Phi=51.8-2.3=49.5 \mu as\), blue dashed line in \ref{Fig6a} and \ref{Fig6b}). 

\begin{figure}[t!]
\begin{center}
\subfloat[Figure illustrating dependence of $\Delta C$ on $r_2$ and $a$ assuming an 
inclination angle of $46^{\circ}$ corresponding to Sgr A*.]
{\includegraphics[scale=0.267]{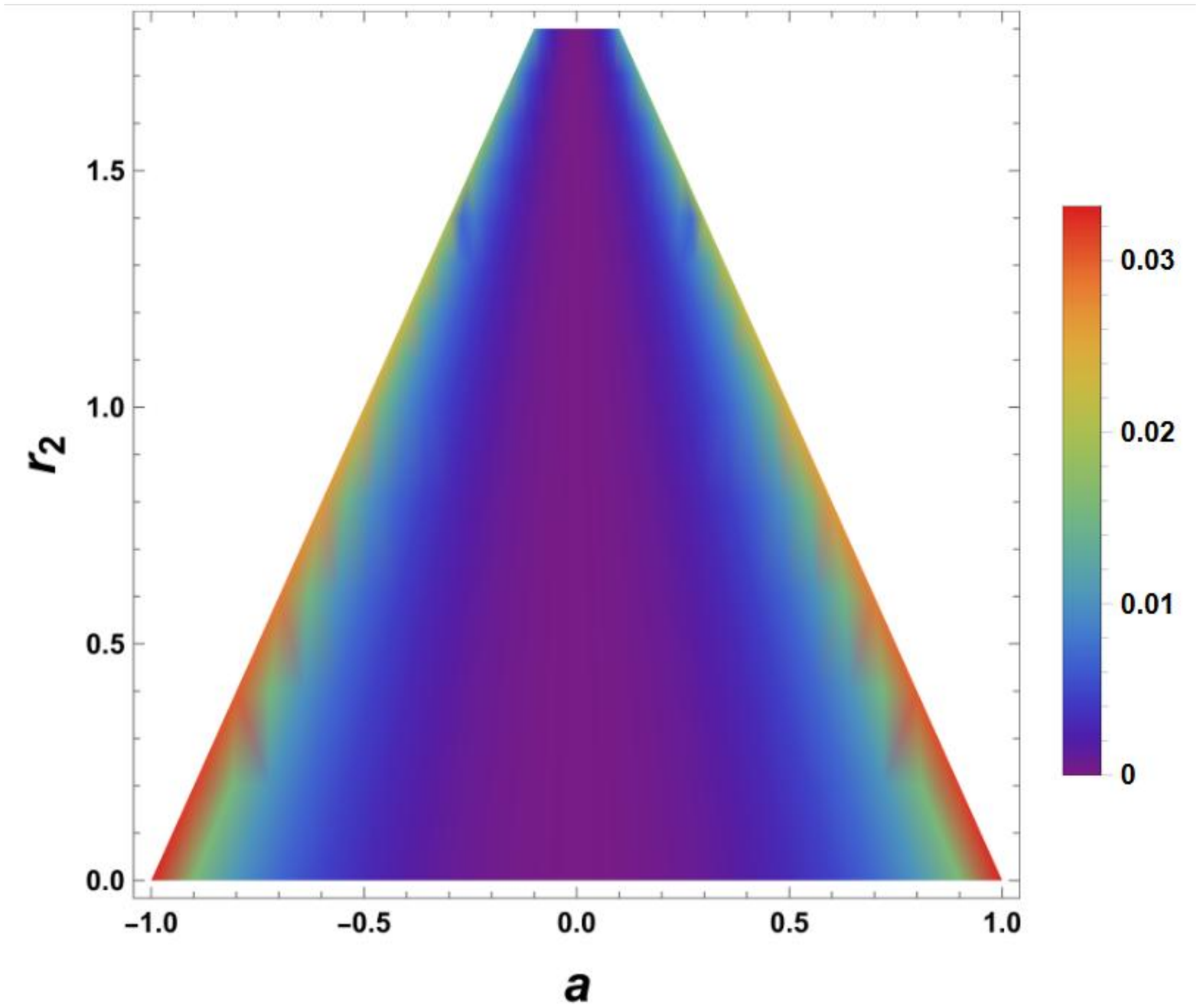}}
\hspace{0.65cm}
\subfloat[Figure illustrating dependence of $\Delta A$ on $r_2$ and $a$ assuming an 
inclination angle of $46^{\circ}$ corresponding to Sgr A*.]
{\includegraphics[scale=0.267]{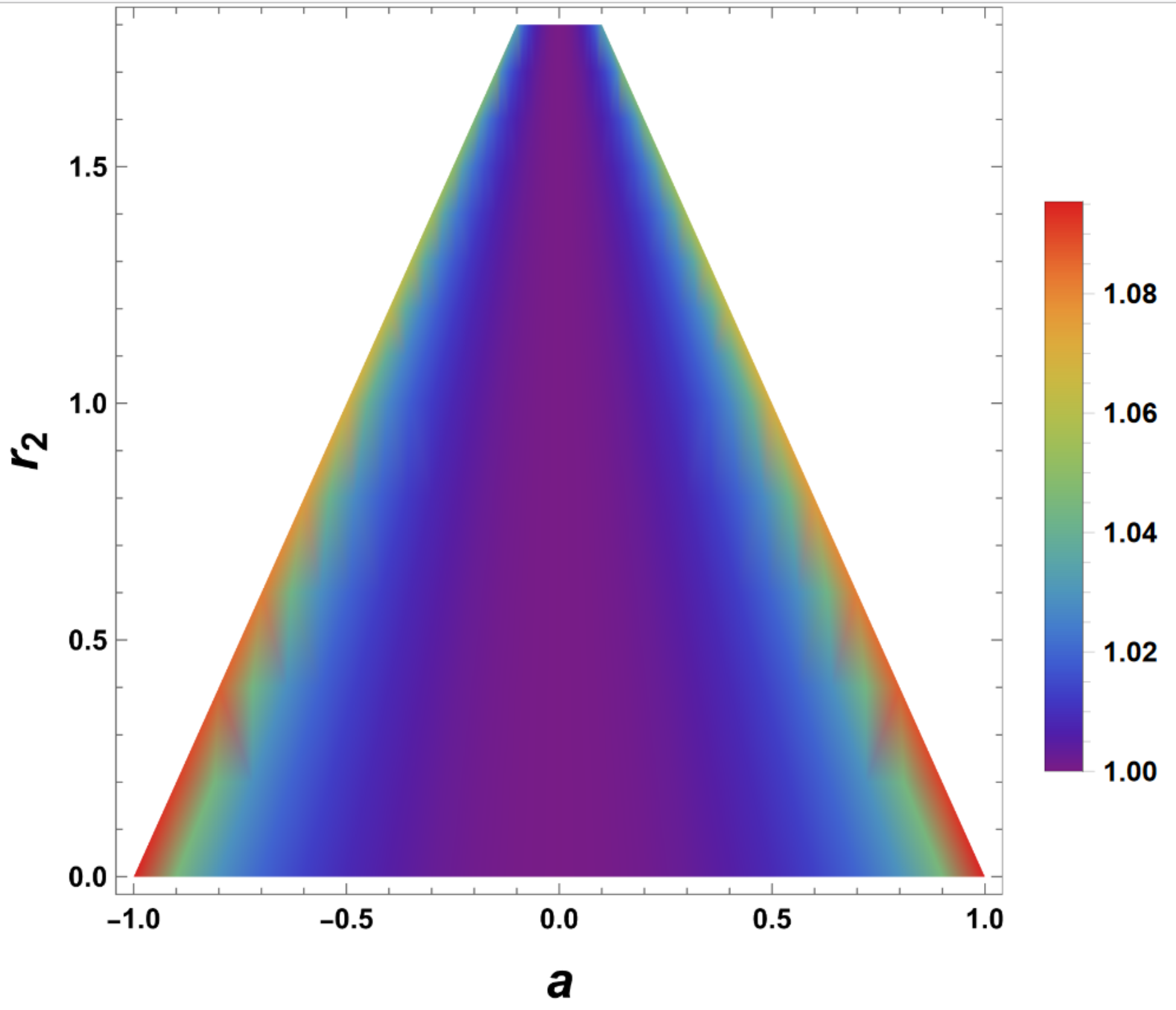}}
\caption{The above figure depicts the variation of $\Delta C$ and $\Delta A$ as function of $r_2$ and $a$, for the source Sgr A*. 
}
\label{Fig7}
\end{center}
\end{figure}

We now discuss the constrains on the dilaton parameter $r_2$ from the distance and mass measurements by the GRAVITY collaboration \cite{GRAVITY:2021xju,GRAVITY:2020gka}. According to the results of the GRAVITY collaboration the mass and distance of Sgr A* turn out to be \(M = (4.261 \pm 0.012) \times 10^{6}M_{\odot}\) and \(D =(8246.7 \pm 9.3)\)pc\cite{GRAVITY:2021xju,GRAVITY:2020gka} respectively. However, when one takes into account the systematics due to optical aberrations, the GRAVITY collaboration constrains the mass and distance of Sgr A* to $M=(4.297 \pm 0.012 \pm 0.040)\times 10^6 M_\odot$ and $D=(8277 \pm 9 \pm 33)$ pc respectively. In \ref{Fig6c} and \ref{Fig6d} we plot contours of theoretical angular diameter of shadow of Sgr A* assuming masses and distances reported by the GRAVITY collaboration. From the figures we note that once again a non-zero dilaton parameter $0.3 \lesssim r_2 \lesssim 0.4$ is required to explain the central value of the observed shadow diameter of \(\Phi=48.7\mu as\) (red solid line in \ref{Fig6c} and \ref{Fig6d}). To reproduce the central value of the image diameter \(\Phi=51.8\mu as\) (blue solid line in \ref{Fig6c} and \ref{Fig6d}), a smaller but non-zero $r_2$ ($0 \lesssim r_2 \lesssim 0.1$) is required. The upper 1-$\sigma$ interval of the image diameter \(\Phi=51.8-2.3=49.5\mu as\), (denoted by the blue dashed line in \ref{Fig6c} and \ref{Fig6d}) can be reproduced by $0.2 \lesssim r_2 \lesssim 0.3$.
Therefore, for all mass and distance estimates of Sgr A*, a small positive dilaton charge is required to explain the observed shadow/image diameter. This implies that the observed shadow of Sgr A* can be better explained by the Kerr-Sen scenario. 

For completeness we plot in \ref{Fig7} the dependence of the the deviation from circularity \(\Delta C\) and the axis ratio \(\Delta A\) on the metric parameters $r_2$ and $a$. This is a theoretical plot which only requires independent estimate of the inclination of the source, which is taken to be $\theta_0=46^\circ$ as discussed earlier. EHT has not provided any data 
related to \(\Delta A\) and \(\Delta C\) for Sgr A*. These results therefore cannot impose additional constrains on the dilaton parameter $r_2$ at present. These plots can however be useful in future when EHT releases data pertaining to \(\Delta A\) and \(\Delta C\) for Sgr A*.






\section{Concluding Remarks}\label{Sec6}
In this work we investigate the signatures of Einstein-Maxwell dilaton-axion (EMDA) gravity in the shadows of Sgr A* and M87* observed by the Event Horizon Telescope collaboration. 
EMDA gravity arises in the low energy effective action of superstring theories and is associated with the dilaton and the axion fields coupled to the Maxwell field and the metric. 
Exploring the astrophysical implications of such a theory is important as it can potentially provide a possibility to test string inspired models. Moreover, the axion and dilaton fields are often invoked to address the inflationary paradigm or the present accelerated expansion of the universe \cite{Sonner:2006yn,Catena:2007jf}.

EMDA gravity is a scalar-vector-tensor theory of gravity that differs substantially from the standard general relativistic scenario. The stationary and axisymmetric black hole solution of EMDA gravity corresponds to the Kerr-Sen spacetime associated with the dilatonic charge and angular momentum acquired from the axionic field. 
The Maxwell field imparts electric charge to the Kerr-Sen black hole and in the absence of the Maxwell field the field strengths corresponding to dilaton and axion vanish. In that event the metric reduces to the Kerr background in \gr. It is important to note that the charge of Kerr-Sen black hole originates from the axion-photon coupling and not from the charged particles falling onto the black hole. 

In the present work we aim to constrain the charge of the Kerr-Sen black hole from observations related to black hole shadow. For this purpose we examine the motion of photons in the Kerr-Sen background and analyze the nature of the light rings and spherical photon orbits. These light rings when projected onto the observer's sky give rise to the black hole shadow which depends sensitively on the background metric. The $x$ and $y$ coordinates of the shadow are dependent on the inclination angle $\theta_0$ and the two impact parameters $\xi$ and $\eta$ which denote the distances of the photon from the axis of rotation and the equatorial plane, respectively. These impact parameters in turn depend on the charge $r_2$, the spin $a$ of the black hole and the radius of the spherical photon orbit $r$. Thus, we have $x$ and $y$ as functions of $\theta_0$, $a$, $r_2$ and $r$. The outline of the shadow $y(x)$ is obtained by eliminating $r$ between $x$ and $y$ which is often achieved numerically. Since $x$ and $y$ also depend on $\theta_0$, $a$ and $r_2$, the shape and size of the shadow depend sensitively on these three parameters. For example, a rapidly rotating black hole viewed at a high inclination angle casts a non-circular shadow. In fact to observe a deviation from circularity in the shape of the shadow, one needs to have non-zero spin and inclination angle, both. The dilaton parameter $r_2$ on the other hand mainly has an impact on the shadow size. An increase in $r_2$ leads to a decrease in the shadow diameter. 

Once the role of the metric on the shadow structure is derived we next compute the various observables associated with the black hole shadow. These include the angular diameter $\Delta \theta$, the axis ratio $\Delta A$ and the deviation from circularity $\Delta C$. Since the shadow is non-circular in general we can define a major axis and a minor axis associated with the shadow and a ratio of the two gives us the axis ratio $\Delta A$. In order to theoretically compute these observables one need to provide estimates of the inclination angle of the source. Computation of $\Delta \theta$ further requires independent estimates of mass and distance of the compact object. For M87* and Sgr A* the mass, distance and inclination angle have been previously determined. 
We use these data to theoretically compute the observables pertaining to the shadow of M87* and Sgr A* which are eventually compared with observations to establish constrains on the dilaton parameter $r_2$. 

When the shadow of M87* is computed with predetermined mass $M=3.5\times 10^9 M_\odot$, obtained from gas dynamics studies, the observed angular diameter cannot be reproduced for any value of $r_2$, including $r_2=0$ which corresponds to the Kerr scenario. This mass measurement therefore possibly needs to be revisited. When mass determined from stellar dynamics measurements ($M=6.2\times 10^9 M_\odot$) is used, the observed shadow diameter can be reproduced by $0.1\lesssim r_2 \lesssim 0.3$ within 1-$\sigma$ only when the maximum offset of 10\% is considered. In both cases the distance is taken to be $D=16.8$ Mpc (obtained from stellar population measurements) to compute the theoretical shadow diameter. For completeness we also compute the shadow diameter with $M=6.5\times 10^9 M_\odot$ which is the mass reported by the EHT collaboration from the shadow measurements assuming GR. Since this mass is larger than the previous measurements, it can explain the observations better. However, theoretical shadow calculated using this mass should not be used to constrain $r_2$ as such a mass is derived from the observed shadow angular diameter assuming GR. With this mass therefore we cannot constrain another alternate gravity model.
We note that with previously estimated masses the theoretical shadow is smaller than the observed one and since an increase in $r_2$ further shrinks the shadow, therefore the Kerr scenario can better explain the image of M87* compared to the Kerr-Sen scenario. It may be worthwhile to mention here that the Kerr solution is not unique to GR but arises even in several other alternative gravity scenarios. For example, the stationary, axisymmetric uncharged black hole solution in string theory resembles the Kerr solution in GR \cite{Thorne:1986iy,Psaltis:2007cw}.
We further mention that the Kerr scenario also fails to reproduce the observed shadow diameter of $(42\pm 3) \mu as$. Therefore, it seems that if in an alternate gravity model the shadow is larger than the GR scenario, it will better explain the observations, e.g. the braneworld scenario \cite{Banerjee:2019nnj}. This may also be the reason why the mass of M87* obtained by the EHT collaboration is larger than the previous two measurements.

For Sgr A*, the mass and distance have been estimated independently by the Keck team and the GRAVITY collaboration. When the theoretical shadow is computed assuming mass and distance estimates by the Keck team, $0.1\lesssim r_2 \lesssim 0.2$ is required to explain the central value of the observed shadow diameter of $\Phi=48.7 \mu as$. Further, $0\lesssim r_2 \lesssim 0.1$ is required to address the observed image diameter ($51.8-2.3=49.5\mu as$) within 1-$\sigma$.
On considering mass and distance estimates by the GRAVITY collaboration, $0.3\lesssim r_2 \lesssim 0.4$ is required to reproduce the central value of the observed shadow diameter while $0\lesssim r_2 \lesssim 0.1$ is necessary to explain the central value of the observed image diameter. Thus, a small positive dilaton charge is required to address the observed shadow of Sgr A* for both the mass and distance estimates by the Keck team and the GRAVITY collaboration. 
The general relativistic scenario however can explain the observations when 1-$\sigma$ interval is considered in the observed shadow diameter. Thus, the Kerr-Sen scenario better explains the observed shadow of Sgr A* compared to the Kerr scenario.

The charge of Kerr-Sen black hole has been constrained previously from different astrophysical observations, e.g. black hole continuum spectrum \cite{Banerjee:2020qmi} and relativistic jets \cite{Banerjee:2020ubc}. Comparison of the theoretical spectrum of eighty Palomar Green quasars with their optical observations reveal that $r_2 \sim 0.2$ best explains the observations. The general relativistic scenario with $r_2=0$ is however included when 1-$\sigma$ interval is considered \cite{Banerjee:2020qmi}. When the jet power associated with ballistic jets in microquasars is used to constrain the dilaton parameter, $r_2\simeq 0$ seem to be favored by observations \cite{Banerjee:2020ubc}. Thus, we note that astrophysical observations, e.g, shadows, continuum spectra or jets either indicate a small dilatonic charge in black holes or exhibit a preference towards the Kerr scenario. All the observations rule out black holes with large dilaton charge. This is interesting as different astrophysical observations on completely different observational samples consistently indicate the same result. The scope to verify this finding will further increase as EHT releases more black hole images with greater resolution.  
The present astrophysical observations like quasi-periodic oscillations observed in the power spectrum of black holes or the Fe-line observed in the black hole spectrum can be further used to verify our result. This will be addressed in a future work.




 
\section*{Acknowledgements}
Research of I.B. is funded by the Start-Up
Research Grant from SERB, DST, Government of India
(Reg. No. SRG/2021/000418).

\bibliography{accretion,KN-ED,regularBh2,Brane,Black_Hole_Shadow,EMDA-Jet,Gravity_1_full,Gravity_2_full,Gravity_3_partial,axion,Reg}

\bibliographystyle{./utphys1}
\end{document}